\documentstyle[12pt,epsf,harvard]{article}
\citationstyle{dcu}

\title{Decoherence limits to quantum computation using trapped ions}
\author{Martin B. Plenio and Peter L. Knight
    \\ Blackett Laboratory,
    Imperial College, London SW7 2BZ, U.K.}

\date{\today}

\begin{document}
\maketitle

\begin{abstract}
We investigate the problem of factorization of large numbers on a
quantum computer which we imagine to be realized within a linear ion
trap. We derive upper
bounds on the size of the numbers that can be factorized on such a
quantum
computer. These upper bounds are independent of the power of the
applied
laser. We investigate two possible ways to implement qubits, in
metastable
optical transitions and in Zeeman sublevels of a stable ground state,
and show that in both cases the numbers that can be factorized
are not large enough to be of practical interest. We also investigate
the effect of quantum error correction on our estimates and show that
in realistic systems the impact of quantum error correction is much
smaller
than expected. Again no number of practical interest can be
factorized.
\end{abstract}

\section{Introduction}
Since Shor's discovery \cite{Shor1,Ekert1} of an algorithm that
allows
the factorization of a large number by a quantum computer in
polynomial
time instead of an exponential time as in classical computing,
interest in the practical realization of a quantum computer has been
much enhanced \cite{Plenio0}. Recent advances in the preparation and
manipulation of
single ions as well as the engineering of pre-selected cavity light
fields have made quantum optics that field of physics which promises
the first experimental realization of a quantum computer. Several
proposals (reviewed in \cite{Barenco0}) for possible experimental
implementations have been made
relying on nuclear spins, quantum dots \cite{Barenco1}, cavity QED
\cite{Sleator1} and on ions in linear traps \cite{Cirac1}.

The realization of a quantum computer in a linear trap in particular
was
regarded as very promising as it was thought that decoherence could
be
suppressed sufficiently to preserve the superpositions necessary for
quantum computation. Indeed, a single quantum gate in such an ion
trap was recently realized by Monroe et al \cite{Monroe1}.
Nevertheless,
the error rate in this experiment was too high to allow the
realization of extended quantum networks. This experiment was not
solely
limited by fundamental processes but rather by technical difficulties
and
one aim of future experiments is to reduce these technical problems
to come closer to the fundamental limits, such that at least small
networks
could be realized.

However, there remains the question whether overcoming technical
problems
will be sufficient to realize practically useful computations such
as factorization of big numbers on a quantum computer in a linear ion
trap. In fact one has to investigate more thoroughly those
limitations
that are due to processes of more fundamental nature than for example
laser linewidths, pulse lengths etc. General considerations
revealed that decoherence will lead to an exponential decrease in the
probability to obtain the correct result of a calculation
\cite{Palma1}.
However, these considerations did not state how fast this exponential
decay
would be and therefore left open the question to whether
factorization will
be possible or not. An analysis of this problem was first carried
out in refs. \cite{Plenio1,Plenio2} where the impact of spontaneous
emission
on the problem of factorization of a large number by a quantum
computer
was considered. It was shown that for realistic laser powers,
factorization
is limited to small numbers (at least if no sophisticated error
correction
methods are implemented). Subsequent investigations have appeared to
be
less pessimistic,
but we note that these did not consider spontaneous emission
but instead concentrated on other effects such as phonon decoherence
\cite{Garg1} or the influence of the ion separation \cite{Hughes1}.

In refs. \cite{Plenio1,Plenio2} limits were obtained that still
depended on
the power of the applied lasers and which therefore leave open the
possibility of improving the maximal bitsize $L$ of the largest
factorizable
number by using high power lasers. Extremely high intensities,
however,
will lead to ionization or, more importantly, to a breakdown of the
two-level-approximation as was mentioned briefly in
\cite{Plenio1,Plenio2}.
In Section II we will use the breakdown of the
two-level-approximation
to derive new estimates for the bitsize $L$ of the factorizable
number.
The new feature in these estimates is their {\em independence}
on the power of the applied laser. We derive intensity independent
limits
of $L$ not only for qubits stored in a metastable optical transition
(see Fig. \ref{Fig1}) but also for qubits stored in the Zeeman
sublevels
of an ion which are then manipulated using
detuned Raman transitions (see Fig. \ref{Fig2}). On the basis of
experimental
parameters for several ions we give estimates for $L_{max}$ and we
show
that spontaneous emission already imposes strong limitations to $L$.
We
then conclude that factorization without the use of efficient error
correction methods is limited to almost trivial numbers if
spontaneous
emission is present, which in reality is inevitable. In Section III
we
therefore proceed to investigate the degree to which such quantum
error
correction methods are able to improve
on these results. It is generally believed that quantum error
correction
is able to increase the number of possible operations substantially,
e.g.
a single error correcting code should allow to perform a number of
operations which is approximately the square of the number of allowed
operations without error correction. We analyze critically this idea
in real atomic systems, especially taking into account the fact that
due to the breakdown of the two-level approximation spontaneous
transitions may leak population out of those levels that represent
the
qubit in the ion. It will be difficult, if not impossible, to bring
this population back into the qubit. This means that after such an
event the quantum error correction code will fail to reconstruct the
state, as the reconstruction procedure works only if the entire
population remains in the qubit states. We take this possibility into
account and obtain estimates that show that the efficiency of
quantum error correction is smaller than expected, although the
application of quantum error correction allows factorization of
somewhat
larger numbers than without the use of quantum error correction. In
Section IV we will then discuss the prospects of factorization and
other
applications of quantum computers in view of the results of the
present
paper. We conclude that spontaneous emission considerations preclude
many applications within the present model of quantum computation.
\section{Decoherence limits without error correction}
\subsection{The linear ion-trap model of quantum computation}
Before we come to the derivation of limits imposed on quantum
computation
by spontaneous emission we briefly describe the ion trap
implementation
of the quantum computer \cite{Cirac1} on which we base our
considerations.
Several ions of mass $M$ are trapped in a linear ion trap such that
they
are lined up and are well separated, ie the next neighbour distance
is
many wavelengths of the lasers used to manipulate the ions. This is
necessary
to be able to address ions separately and can lead to further
restrictions
on quantum computation in this model \cite{Hughes1,James1}. For a
schematic
 picture see Fig. \ref{Fig3}. The motional degrees
of freedom of the ions, and especially their collective
center-of-mass
(COM) motion with frequency $\nu$, are cooled to the ground state. In
order to be able to implement two-bit gates in this scheme we
use the COM mode as a bus which allows us to create entanglement
between
different ions. This is achieved with an interaction that creates a
phonon
in the COM mode when we deexcite an ion and annihilates a phonon in
the COM
mode when an ion is excited. This interaction is generated by a laser
which
is detuned from the transition frequency by $\Delta=-\nu$. In that
case the
Hamilton operator in the Lamb-Dicke limit, the RWA and a suitable
interaction
picture is given by \cite{Cirac1}
\begin{equation}
	H = \frac{\eta}{\sqrt{5L}}\,\frac{\Omega_{01}}{2}
	\left[ |1\rangle\langle 0| a +
	|0\rangle\langle 1|a^{\dagger}\right]\; ,
	\label{1}
\end{equation}
where $\eta=\frac{2\pi}{\lambda}\sqrt{\frac{\hbar}{2 M \nu}}$ is the
Lamb-Dicke parameter, $\Omega_{01}$ is the Rabi frequency on the
$0\leftrightarrow 1$ transition where levels $0$ and $1$ represent
the
corresponding logical values of the qubit. The denominator
$\sqrt{5L}$
originates from the fact that we are considering the COM mode which
has
an effective mass $5LM$ because all $5L$ ions are oscillating in the
trap
potential. The trap
has to contain $5L$ ions, as this is the number of ions that is
required
to implement Shor's algorithm to factorize an $L$ bit number
\cite{Shor1,Vedral1} (actually the correct number is $5L+2$ but
for simplicity we drop the $2$). It is further assumed that coupling
to other vibrational levels can be neglected. This assumption is a
reasonable first approximation as the closest lying vibrational mode
has frequency $\sqrt{3}\nu$ independent of the number of ions in the
trap. Nevertheless, this approximation breaks down when the Rabi
frequency becomes too large, ie. if $\Omega_{01}\approx \nu$
\cite{Hughes1,James1}. For our
considerations of the influence of spontaneous decay we neglect this
effect although it may well become important in longer calculations.
Using the Hamilton operator eq. (\ref{1}) it can be shown that it is
possible to construct a CNOT gate with only four $\pi$-rotations, and
to construct the more involved Toffoli gate \cite{Barenco3} with six
$\pi$-rotations \cite{Cirac1}.
In the following we will define an elementary time step in terms of
the computation time that is required to perform a CNOT gate. All
other
performance times can be reexpressed in units ofthat of the CNOT
gate.
The implementation of Shor's algorithm also requires a number of one
bit operations, ie. operations that leave the COM mode unaffected
during their performance. However, these operations can be performed
much faster than two-bit gates, because $\frac{\eta}{\sqrt{5L}}\ll 1$
and also because fewer individual laser pulses are required.
Therefore
their contribution to calculation time and decoherence is small and
will be neglected in the following.
\subsection{Qubits stored in two-level systems}
After this short discussion of the
ion trap model we will now derive the intensity independent estimates
for $L$, first for a qubit which is stored in a metastable two-level
transition (see Fig. \ref{Fig1}). Later in this section we will also
consider
the case of a qubit stored in Zeeman sublevels manipulated by
strongly
detuned Raman pulses (see Fig. \ref{Fig2}). First we assume that the
two-level approximation holds and derive an expression for the
computation
time that is required to factorize an $L$ bit number. As the next
step we then include in our calculation possible extraneous levels
and
their spontaneous decay.

As stated above, a CNOT gate can be implemented by four
$\pi$-pulses according to the Hamilton operator eq. (\ref{1}). This
requires an elapsed time
\begin{equation}
	\tau_{el} = 4\,\frac{\pi\sqrt{5L}}{\eta\Omega_{01}}\; ,
	\label{2}
\end{equation}
which we will call the elementary time step. It is known
\cite{Vedral1}
that $\epsilon L^3 + O(L^2)$ of these elementary time steps are
required to
implement Shor's algorithm using elementary gates such as one bit
gates, CNOT
gates, Toffoli gates and Fredkin gates. Therefore the total
computation
time to complete a factorization will be
\begin{equation}
	T \cong \frac{4\,\pi\sqrt{5L}}{\eta\Omega_{01}}\epsilon L^3\; ,
	\label{3}
\end{equation}
which assumes that the gates are executed one after the other with
zero
time delay between two gates. The value of the constant $\epsilon$
very much depends on the actual practical implementation of Shor's
algorithm. The implementation given in \cite{Vedral1} requires
$80\, L^3$ CNOT gates, $80\, L^3$ Toffoli gates and $8\, L^3$ Fredkin
which result in $\epsilon = 80 + 1.5\cdot 80 + 2\cdot 80 =216$. We
neglect here all contributions of order $L^2$ which give significant
corrections only for small $L$.

If the quantum computation is to give a useful answer,
no spontaneous emission is allowed to occur during the whole
calculation
time $T$, because an emission usually alters the wavefunction of the
quantum computer completely. We will illustrate this point for a
quantum computer which performs a discrete Fourier transformation
(DFT)
\cite{Barenco0}. Qualitatively similar results were obtained in ref.
\cite{Barenco2} for phase errors.

Spontaneous emission in fact leads to two sources of errors. One,
quite
obviously, arises from the actual spontaneous emissions in the
quantum
computer. The other one is less obvious and is due to the conditional
time evolution when no spontaneous emission has taken place. This
time
evolution differs from the unit operation because the failure to
detect
a photon provides us with information about the system which is
reflected
in a change of the wavefunction \cite{Plenio4}.

First we illustrate the case of an unstable quantum computer which
performs a DFT on a function which is evaluated at $32$ points. The
resulting square modulus of the wavefunction of the quantum computer
is compared to the exact result obtained from a absolutely stable
quantum computer. The function on which we perform the DFT is given
by
$f(n)=\delta_{8,(n\,mod  10)}$ for $n=0,1,\ldots,31$. We have
implemented
the Hamilton operators (in the Lamb-Dicke limit) for all the
necessary
quantum gates in a linear ion trap \cite{Cirac1} to realize this DFT.
In
addition to the coherent time evolution we also take into
account possible spontaneous emissions from the upper levels of the
ions
but we neglect all other sources of loss. To calculate the time
evolution
of the quantum computer, we then use the quantum jump approach
\cite{Plenio4,Dalibard1,Carmichael1,Knight1,Hegerfeldt1}
to simulate the time evolution of a {\em single}
quantum computer. The simulation runs as follows: We generate a
random
number and compare this random number after each time step with the
squared norm of the wavefunction of the quantum computer (The norm of
the wavefunction decreases as it represents the probability that no
photon has been emitted). If the squared norm of the wavefunction is
smaller than the random number, an emission is deemed to have
occured.
We then generate another random number and continue our simulation.
Two
results of our simulations are shown  in Figs. \ref{Fig5} and
\ref{Fig6}.
In Fig. \ref{Fig5} one emission has taken place during the
calculation time
of the quantum computer. If we compare the resulting wavefunction
with the
correct wavefunction, we observe a marked difference between the two.
In Fig. \ref{Fig6} we show the wavefunction of an unstable quantum
computer which has not suffered a spontaneous emission during the
calculation of the DFT. We clearly see that even when no spontaneous
emission has taken place, the wavefunction of the quantum computer
differs
substantially from the correct result. This difference becomes
stronger
and stronger the larger the ratio between the computation time $T$
and the spontaneous lifetime $\tau_{sp}$ of the quantum computer
becomes.
Therefore the wavefunction of the quantum computer will be
sufficiently
close to the correct result only if the whole computation is finished
in a time $T$ that is much shorter than the spontaneous lifetime
$\tau_{sp}$
of the quantum computer.

Therefore we need to determine the
spontaneous lifetime of the quantum computer. It is reasonable to
assume
that each ion has an average excitation of $0.5$ during the whole
calculation.
The reason for this is that the quantum computer is in a
superposition
of very many states each of which represents a string of logical
states $0$
and $1$. The probability to find an atom excited will on average be
$0.5$.
In addition during the computation each ion will suffer many $2\pi$
rotations
which again leads to an average population of $0.5$. As the ions are
separated by many wavelengths it is reasonable to assume that each
ion
decays independently of all others with a decay rate
of $2\Gamma_{11}$. As the quantum computer consists of $5L$ ions the
spontaneous lifetime of our quantum computer is
\begin{equation}
	\tau_{sp} = \frac{1}{5L \Gamma_{11}} \;\; .
	\label{4}
\end{equation}
Therefore we obtain the condition
\begin{equation}
	p_{em}^{(1)} = \frac{T}{\tau_{sp}} \ll 1\;\; ,
	\label{5}
\end{equation}
and from that
\begin{equation}
	\frac{\Omega_{01}}{\Gamma_{11}} =
	\frac{20\pi\epsilon \sqrt{5L^9}}{\eta p_{em}^{(1)}}\;\; .
	\label{6}
\end{equation}
We can now insert eq. (\ref{6}) into eq. (\ref{3}) and obtain
\begin{equation}
	T = \frac{400\pi^2 \epsilon^2}{\eta^2 p_{em}^{(1)}}
	\frac{\Gamma_{11}}{\Omega_{01}^2} L^8
	\; ,
	\label{7}
\end{equation}
where we have isolated the fraction $\Gamma_{11}/\Omega_{01}^2$
because it can be expressed in a form that is independent of the
actual
type of transition we are considering (electric dipole,
quadrupole,...).
We find
\begin{equation}
	\frac{\Omega_{01}^2}{\Gamma_{11}} = \frac{6\pi c^3 \epsilon_0}
	{\hbar\omega^3_{01}} \, E^2\;\; ,
	\label{8}
\end{equation}
which can be derived along similar lines as in the usual quantum
mechanical
derivation of the relation between the Einstein $A$ and $B$
coefficient.
If we want to reduce the estimated computation time $T$ we have to
have
a ratio $\frac{\Omega_{01}^2}{\Gamma_{11}}$ which is as large as
possible.
As the transition frequency is given by the ion we can adjust only
the field strength of the laser. However, there is an upper limit to
this
field strength which is approximately given by the field strength
between
an electron and the proton in a hydrogen atom in its ground state.
The value is
\begin{equation}
	E_{hyd} = \frac{e}{4\pi\epsilon_0 a_0^2} =
	5.52\, 10^{11} \frac{V}{m}\;\; ,
	\label{9}
\end{equation}
where $a_0=\frac{4\pi\epsilon_0\hbar^2}{e^2 m_0}$. This would yield a
maximum value of $1.46\,10^{24}$ for eq. (\ref{8}). However, one
should
note that this field strength is sufficient to destroy the ion within
one optical cycle by tunnel ionization \cite{Augst1} and represents
therefore a very hypothetical upper
limit. In fact we will see later in this section that, not
surprisingly,
other approximations such as the two-level approximation break down
much
earlier.

We may use eq. (\ref{7}) to obtain a first estimate for an upper
limit on
$L$ by requiring that $T$ has to be smaller than the decoherence time
$\tau_{aux}$ due to all other decohering processes that may occur,
such as
collisions, stray fields, phonon losses, laser phase fluctuations,
\ldots.
Nevertheless, from a fundamental point of view the situation is not
yet
entirely satisfying, as in principle higher laser powers and
technical
improvements concerning the auxiliary decoherence effects could
change
the estimates quite drastically. Therefore we will now proceed to
show
that the inclusion of spontaneous emission from auxiliary levels into
the dynamics will lead to an estimate for the upper limit of $L$
which
is entirely {\em intensity independent}.

To model the influence of extraneous levels on the decoherence rate
due
to spontaneous emissions we assume that there is one other level $2$
present which couples to both levels $0$ and $1$ by the same laser
that drives the $0\leftrightarrow 1$ transition. As the contribution
of even
further detuned levels decreases rapidly with the detuning (it can be
shown
that the infinitely many states in the atom give only a finite
contribution)
we restrict our treatment to one additional level. As the laser is
strongly
detuned from the $i\leftrightarrow 2$ transitions the population in
level
$2$ will be very small and we can safely assume that the
contributions
from levels $0$ and $1$ to the population of level $2$ add up. The
population
in level $2$ will then be
\begin{equation}
	\rho_{22} = \frac{1}{2}
	\left\{ \frac{\Omega_{02,eff}^2}{4\Delta_{02}^2} +
	\frac{\Omega_{12,eff}^2}{4\Delta_{12}^2} \right\}\;\; ,
	\label{12}
\end{equation}
where $\Omega_{i2,eff}$ is the Rabi frequency on the
$i\leftrightarrow 2$
transition and $\Delta_{i2}$ the detuning on that transition. In
eq. (\ref{12}) we have not taken into account that for very large
detuning
the RWA  is not very good any more. However, the contribution of all
possible atomic levels should be approximated well by eq. (\ref{12}).
We have used
the notation $\Omega_{12,eff}$ because the Rabi frequency on the
$i\leftrightarrow 2$ transition now depends on the type of the
$i\leftrightarrow 0$ transition. There are two cases.
\begin{description}
\item{a)} If the $0\leftrightarrow 1$ transition is an electric
quadrupole transition (E2) then for the excitation of phonons it is
necessary to place the ion at the antinode of the electric field of
the laser. Then the same laser on an electric dipole transition (E1)
will, in leading order, leave the COM mode untouched. This implies
that
$\Omega_{i2,eff}=\Omega_{i2}$
\item{b)} If the $0\leftrightarrow 1$ transition is an electric
octupole transition (E3) then for the excitation of phonons it is
necessary
to place the ion at the node of the electric field of the laser.
Then the same laser on an electric dipole transition (E1) will also
excite phonons in the COM mode. This implies that
$\Omega_{i2,eff}=\frac{\eta'\Omega_{i2}}{\sqrt{5L}}$ where $\eta'$
is the Lamb-Dicke parameter on the $i\leftrightarrow 2$ transition.
\end{description}
We will have to distinguish between the two possibilities a) and b),
the first of which occurs for example in
$\mbox{Ba}^{+}$,$\mbox{Ca}^{+}$
and $\mbox{Hg}^{+}$ while the second one occurs in $\mbox{Yb}^{+}$
(an ion which might allow factorization of very large numbers
according
to estimates which do not include spontaneous emission
\cite{Hughes1}) .
We first deal with case a) and then state the result for case b).
In case a) we obtain
\begin{equation}
	\rho_{22} = \frac{1}{2}
	\left\{ \frac{\Omega_{02}^2}{4\Delta_{02}^2} +
	\frac{\Omega_{12}^2}{4\Delta_{12}^2} \right\}\;\; .
	\label{13}
\end{equation}
As in the pure two-level case, we want to be sure that no spontaneous
emission will take place during the whole calculation, neither from
level
$1$ nor from level $2$. This implies the condition
\begin{equation}
	2\Gamma_{22}\rho_{22} T = p_{em}^{(2)} \ll 1\;\; .
	\label{14}
\end{equation}
We now use eq. (\ref{7}) and
\begin{equation}
	\frac{\Gamma_{11\rightarrow 00}}{\Omega_{10}^2}
	\frac{\Omega_{02}^2}{\Gamma_{22\rightarrow 00}} =
	\left( \frac{\omega_{10}}{\omega_{21}} \right)^3
	\label{15}
\end{equation}
which derives from eq. (\ref{8}) and where $2\Gamma_{ii\rightarrow
00}$
is the decay rate of level $2$ on the $i\leftrightarrow 0$ transition
and $\omega_{i0}$ the corresponding transition frequency.
Inserting eqs. (\ref{7}) and (\ref{15}) into eq. (\ref{13}) we obtain
\begin{equation}
	L = \left\{ \frac{\eta^2 p_{em}^{(1)} p_{em}^{(2)} }
	{100 \pi^2 \epsilon^2 \Gamma_{22}^2 } \,
	\frac{1}{ \frac{\Gamma_{22\rightarrow 00}}{\Delta_{20}^2
\Gamma_{22}}
	\left(\frac{\omega_{10}}{\omega_{02}}\right)^3 +
	\frac{\Gamma_{22\rightarrow 11}}{\Delta_{21}^2 \Gamma_{22}}
	\left(\frac{\omega_{10}}{\omega_{21}}\right)^3 }
	\right\}^{1/8}\;\; ,
	\label{16}
\end{equation}
where we have assumed that $\Gamma_{11}=\Gamma_{11\rightarrow 00}$,
ie.
level $1$ only decays back into state $0$.
For case b) we obtain a slightly different result which nevertheless
has a very similar structure. We find
\begin{equation}
	L = \left\{ \frac{p_{em}^{(1)} p_{em}^{(2)} }
	{20 \pi^2 \epsilon^2 \Gamma_{22}^2} \,
	\frac{1}{\frac{\Gamma_{22\rightarrow 00}}{\Delta_{20}^2 \Gamma_{22}}
	\left(\frac{\omega_{10}}{\omega_{02}}\right) +
	\frac{\Gamma_{22\rightarrow 11}}{\Delta_{21}^2 \Gamma_{22}}
	\left(\frac{\omega_{10}}{\omega_{21}}\right) }
	\right\}^{1/7}\;\; ,
	\label{17}
\end{equation}
where we have used the fact that the Lamb-Dicke parameter depends on
the transition frequency so that
\begin{equation}
	\frac{\eta_{10}}{\eta_{20}} = \frac{\omega_{10}}{\omega_{02}}
	\;\; .
	\label{18}
\end{equation}
It is important to note that the expressions eqs. (\ref{16}) and
(\ref{17})
are {\em independent} of the laser power as long as
$\Omega_{i2}\ll\Delta_{i2}$.
This independence has its origin in the fact that increasing the
laser
power gives rise to two competing effects. It decreases the
computation
time $T$ given in eq. (\ref{7}) and therefore decreases the
probability
for an emission from level $2$. On the other hand an increased laser
power
increases the population in level $2$ which increases the probability
for
a spontaneous emission. Both effects cancel, leading to the intensity
independent results eqs. (\ref{16}) and (\ref{17}).

In table 1 we give values for the bounds eq. (\ref{16}) and
(\ref{17})
for realistic ions. The values for $Ba, Hg$ and $Ca$ were calculated
using eq. (\ref{16}) as the qubit transition is quadrupole allowed,
while $Yb$
with an octupole allowed qubit transition is an example of case b)
and
is therefore calculated according to eq. (\ref{17}). We calculate the
values
for $L$ assuming $p_{em}^{(1)}=p_{em}^{(2)}=1$, which is the most
optimistic
choice, and for two values of the
Lamb-Dicke parameter: an optimistic $\eta=1$ and the more realistic
$\eta=0.01$ \cite{James1}. We see that the numbers that may be
factorized
on a quantum computer using metastable optical transitions are very
small
even for the optimistic choice $\eta=1$. Only the $Yb$ ion gives
results
from which one may hope to factorize at least small numbers. However,
it
should be realized that it is extremely difficult to drive the
hyperstable
qubit transition in $Yb$ sufficiently quickly to finish the
calculation
in a reasonable time. If we assume a ratio
$\Omega_{01}^2/\Gamma_{11}=10^{16}$ where $\Gamma_{11}=3.77\cdot
10^{-9}$
then from eq. (\ref{3}) we find for $\eta=1$ the value $T=126s$ for a
$4$
bit number. This is so long that it is very unlikely that one can
isolate
the system during that time from all other decoherence sources.
Therefore
the practical limit to $L$ for $Yb$ is probably smaller than the one
given
in table 1.
\subsection{Qubits as Zeeman sublevels of stable ground state}
So far we have dealt with the case where the qubit is stored in a
metastable optical transition. This method certainly has the
disadvantage
that at practically all times the quantum computer has a mean
excitation
of about $0.5$ per qubit. Therefore, even if a qubit does not take
part in a
quantum gate operation, the qubit will decohere due to spontaneous
emissions
from the excited state of the qubit. To circumvent this problem one
would like
to store the qubits in Zeeman sublevels of a stable ground state. In
that
case stored qubits do not suffer any decoherence due to spontaneous
emissions. However, how do we manipulate the qubit? It is
unfortunately
not practical to drive the qubit with a microwave field directly
because
the very long wavelength of this radiation does not allow us to
address single
ions as they cannot be separated far enough in a linear trap.
Therefore we
have to use a different method of manipulating the qubit. This method
uses
the presence of another level $2$ (see Fig. \ref{Fig2} ) which is
coupled
to the two
qubit levels $0$ and $1$ by two lasers of Rabi frequency
$\Omega_{02}$
and $\Omega_{12}$. Each laser is strongly detuned from its
transition,
and its Rabi frequency is much smaller than the detuning.
However, it is assumed that both detunings are equal, ie. the two
photon
detuning vanishes while the one photon detuning $\Delta_2$ is large.
The
advantage of this method is that due to the strong detuning
$\Delta_2$ the
population in level $2$ is small, so that spontaneous emissions from
that
level are rare. A Hamilton operator that implements an interaction
analogous
to the one generated by eq. (\ref{1}), ie the qubit is excited while
a
phonon in the COM mode is deexcited and vice versa, is given by
\begin{equation}
 	H = -\hbar\Delta_2|2\rangle\langle 2| +
	\frac{\hbar\Omega_{02}}{2}
	\left[ |2\rangle\langle 0|  + |0\rangle\langle 2| \right]
	+
	\frac{\hbar\eta\Omega_{12}}{2\sqrt{5L}}
	\left[ |2\rangle\langle 1| a  +
	|1\rangle\langle 2| a^{\dagger}\right]
	\; ,
	\label{19}
\end{equation}
where again the Lamb-Dicke limit, the RWA and a suitable interaction
picture is used. To make sure that the Raman transitions have the
effect as full Rabi oscillations between levels $0$ and $1$ the
condition
\begin{equation}
	\Omega_{02} = \frac{\eta\Omega_{12}}{\sqrt{5L}}
	\label{20}
\end{equation}
has to be satisfied which we will assume in the following.
The condition eq. (\ref{20}) can be obtained from the solution of
the time evolution
\begin{eqnarray}
	|\langle 0|\psi(t)\rangle|^2 &=&
	\left| \frac{ \Omega_{02}^2
	e^{-i(\Omega_{02}^2+\eta^2\Omega_{12}^2/5L)t/4\Delta_2 }
	+ \eta^2\Omega_{12}^2/5L }{\Omega_{02}^2 +\Omega_{12}^2} \right|^2
	\label{21}\\
	|\langle 1|\psi(t)\rangle|^2 &=& 1 -
	|\langle 0|\psi(t)\rangle |^2 \;\; .
	\label{22}\\
	|\langle 2|\psi(t)\rangle|^2 &=& \frac{\Omega_{02}^2}{2\Delta_2^2}
	\left(
	1 + \cos \Delta_2 t \right)
	\label{23}
\end{eqnarray}
which is valid for $\Gamma_{22} t \ll 1$ and for an initial state
$|\psi(t)\rangle = |0\rangle$. To obtain
$|\langle 0|\psi(t)\rangle|^2 = 0$ for some $t$ eq. (\ref{20}) has to
be satisfied. Before we continue we have to realize that it is indeed
enough to consider the time evolution eqs. (\ref{21}-\ref{23}) and
the
corresponding time evolution for the initial condition
$|\psi(t)\rangle = |1\rangle$. To see this we realize that to a good
approximation the coherences between levels $0$ and $1$ are uniformly
distributed. Averaging the initial state over this distribution then
yields a mixed state as an initial state in which both states $0$ and
$1$ have equal weight. Therefore calculating the time evolution for
both possible initial states separately and averaging over the two
results leads to the final result.

Again in the Raman pulse implementation of qubits a CNOT gate can
be implemented with four $\pi$ pulses. From the solution eqs.
(\ref{21}-\ref{23})
we see that the effective Rabi frequency of the
qubit transition is given by
\begin{equation}
	\Omega_{eff} = \frac{\Omega_{02}^2}{2\Delta_2}\;\; .
	\label{24}
\end{equation}
Therefore the performance of a CNOT gate requires the time
\begin{equation}
	\tau_{el} = \frac{8\pi\Delta_2}{\Omega_{02}^2}\;\; .
	\label{25}
\end{equation}
Again the probability for a spontaneous emission from level $2$ has
to be small, ie.
\begin{equation}
	2\Gamma_{22} \rho_{22} \tau_{el} \epsilon L^3 =
 	p_{em}^{(2)} \ll 1\;\; .
	\label{26}
\end{equation}
This condition immediately yields
\begin{equation}
	L = \left( \frac{\Delta_2 p_{em}^{(2)}}
	{8\pi\epsilon\Gamma_{22}} \right)^{1/3}
	\;\; .
	\label{27}
\end{equation}
which is already intensity independent. From eq. (\ref{27}) one could
conclude that one is able to factorize gigantic numbers by making the
ratio $\Delta_2/\Gamma_{22}$ sufficiently large, eg.
$\Delta_2=10^{13}s^{-1},
\Gamma_{22}=1s^{-1}$ and $\epsilon=216$ would lead to $L=1225$.
But again this reasoning is invalid because we have neglected the
fact
that the two-level approximation breaks down because in the ion other
levels, which we model by a level $3$, also couple to levels $0$ and
$1$ via
the Raman pulses. This additional coupling is very important because
it
potentially involves levels that may have $\Gamma_{33}\cong
10^8s^{-1}$.
In addition the presence of level $3$ changes the effective Rabi
frequency
in the qubit and may therefore lead to additional errors. Finally
there
is a limit to the size of $\Delta_2$ from the nature of the qubit
quantum
numbers, as usually states $0$ and $1$ are hyperfine levels which
require
a nuclear spin flip which is precluded for very large $\Delta_2$.

To achieve a more restrictive, but again intensity independent, upper
bound
for $L$ than the one given in eq. (\ref{27}) we now take into account
the presence of one more level $3$ which acquires population during
the
execution of the quantum gate. One can give a complete analysis of
the
full four level system; however, this is very tedious, although not
complicated in principle. In this paper we will restrict ourselves to
two
limiting cases where the analysis is simpler and already reveals the
basic physics. We avoid an intermediate regime which is also not of
practical interest as the precise control of the qubits will be
difficult. If we denote the Rabi frequencies on the $i\leftrightarrow
3$
transition by $\Omega_{i3}$ and the one photon detuning as
$\Delta_{3}$ we can
distinguish two cases, one of which reduces to the analysis given
above,
the other one requiring further investigation. First consider the
case
where
\begin{equation}
	\frac{\Omega_{03}^2}{\Delta_3} \gg \frac{\Omega_{02}^2}{\Delta_2}
	\;\; .
	\label{28}
\end{equation}
In that case the analysis given above, neglecting level $3$, was not
correct
as the influence of level $3$ on the dynamics of the qubit is
actually
larger than that of level $2$. For example the proper effective Rabi
flopping
frequency of the qubit is then given by $\Omega_{03}^2/2\Delta_3$
instead
of $\Omega_{02}^2/2\Delta_2$. Due to relation eq. (\ref{8}) we can
conclude
from eq. (\ref{28}) that
\begin{equation}
	\frac{\Gamma_{33\rightarrow 00}}{\Delta_3}
	\gg \frac{\Gamma_{22\rightarrow 00}}{\Delta_2}
	\label{29}
\end{equation}
and therefore
\begin{equation}
	\left( \frac{\Delta_3 p_{em}^{(3)}}
	{8\pi\epsilon\Gamma_{33}} \right)^{1/3} \ll
	\left( \frac{\Delta_2 p_{em}^{(2)}}
	{8\pi\epsilon\Gamma_{22}} \right)^{1/3}\;\; .
	\label{30}
\end{equation}
In fact what one has to do is to redo the
analysis leading to eq. (\ref{27}) but replacing level $2$ by level
$3$.
The other case is given by
\begin{equation}
	\frac{\Omega_{03}^2}{\Delta_3} \ll \frac{\Omega_{02}^2}{\Delta_2}
	\;\; ,
	\label{31}
\end{equation}
which means that indeed level $2$ determines the dynamics of the
system.
In that case we immediately find
\begin{equation}
	L = \left( \frac{\Delta_2 p_{em}^{(3)}}
	{8\pi\epsilon\Gamma_{22}} \right)^{1/3} \ll
	\left( \frac{\Delta_3 p_{em}^{(2)}}
	{8\pi\epsilon\Gamma_{33}} \right)^{1/3}\;\; ,
	\label{32}
\end{equation}
where $\Delta_{3}$ and $\Gamma_{33}$ is more or less determined
because
the lasers are extremely far detuned from the $i\leftrightarrow 3$
transitions. This could already serve as a strong limit for $L$,
however,
it is of a different structure compared to the intensity independent
limit
eqs. (\ref{16}-\ref{17}) for the two-level case. It is interesting
to note that it is possible to construct a limit which has
a form completely analogous to eqs. (\ref{16}-\ref{17}) only with
slightly
different exponents. For this derivation we assume that level $2$
influences
the qubit dynamics the most, ie. eq. (\ref{31}) is satisfied.
This assumption itself together with $\Omega_2\ll\Delta_2$
(detuned Raman pulses) yields a lower limit to the computation time
\begin{equation}
	T = \frac{8\pi\Delta_2}{\Omega_{02}^2} \epsilon L^3
	\gg \frac{8\pi\epsilon L^3}{\Delta_2}
	\gg \frac{8\pi\epsilon L^3
	\Gamma_{33\rightarrow 00}}{\Gamma_{22\rightarrow 00}\Delta_3}\;\; .
	\label{32a}
\end{equation}
We will see from this expression for the computation time that any
system
with an exceedingly high lifetime of level $2$ is practically
useless for quantum computation because the time required to perform
a
computation is extremely high.

Again of course it is important to distinguish between the cases a)
and b)
for the different types of qubit transitions (electric quadrupole,
octupole,
..)  as we have done when we investigated the case of a two-level
system as a
qubit. We will treat case a) explicitly in the following while we
only
state the result for case b).

To calculate the influence of level $3$ to the decoherence we first
have to
find out the average population in that level. For this let us
realize
that during the implementation of a CNOT a full $4\pi$-rotation will
be
performed. That means that on average the population in both levels
$0$
and $1$ is equal. Therefore the much slower coupling to level $3$
sees an
averaged population in both levels with no average coherence. We
can therefore average over two contributions to $\rho_{33}$ which
originate from the initial states $|0\rangle$ and $|1\rangle$ and we
obtain
\begin{equation}
	\rho_{33} = \frac{1}{2}\,
	\frac{\Omega_{13}^2 + \eta'^2\Omega_{03}^2/5L}{4\Delta_{3}^2}\;\; ,
	\label{33}
\end{equation}
where $\eta'$ is the Lamb-Dicke parameter for the $0\leftrightarrow
3$
transition. EQ. (\ref{33}) contains an additional factor $1/2$
compared to
eq. (\ref{23}). This is due to the fact that level $3$ is strongly
decaying
and that therefore $\Gamma_{33} t\gg 1$. Note also that the roles of
the lasers
on the $i\leftrightarrow 3$ transitions has been reversed compared to
their role
on the $i\leftrightarrow 2$ transitions; the phonon is now destroyed
by an
excitation from level $0$ and not from level $1$ anymore. This is due
to the fact
that $i\leftrightarrow 3$ transitions are dipole allowed (E1) while
the $i\leftrightarrow 2$ transition is quadrupole allowed (E2). Again
an
emission from level $3$ should not occur during the whole calculation
which
means
\begin{eqnarray}
	p_{em}^{(3)} &=& 2\Gamma_{33} \rho_{33} \tau_{el} \epsilon L^3
	\nonumber\\
	&=& \frac{16\pi^2\epsilon^2\Gamma_{22}\Gamma_{33}
	(\Omega_{13}^2 + \eta'^2\Omega_{03}^2/5L)}
	{\Delta_3^2 \Omega_{02}^2 p_{em}^{(2)} } L^6\;\; .
	\label{34}
\end{eqnarray}
Now we have to find out which term in the brackets of eq. (\ref{34})
is dominant. Because we need $\Omega_{02}=\eta\Omega_{12}/\sqrt{5L}$,
see eq. (\ref{20}), and because we are in the Lamb-Dicke limit, ie.
$\eta/\sqrt{5L}\ll 1$, the laser on the $0\leftrightarrow 2$
transition
has to be weaker than the one on the $1\leftrightarrow 2$ transition.
But again $\eta'/\sqrt{5L}\ll 1$ and therefore
$\Omega_{13}\ll\eta'\Omega_{03}/\sqrt{5L}$. This implies
\begin{eqnarray}
	p_{em}^{(3)} &=& \frac{16\pi^2\epsilon^2}{\Delta_3^2 p_{em}^{(2)} }
	\;\frac{\Gamma_{22}\Gamma_{33}\Omega_{13}^2}{\Omega_{02}^2} L^6
	\nonumber\\
	&=& \frac{16\pi^2\epsilon^2\Gamma_{22}\Gamma_{33} }
	{\Delta_3^2 p_{em}^{(2)} } \,
	\frac{\Gamma_{22\rightarrow 00}}{\Omega_{02}^2}\,
	\frac{\Omega_{13}^2}{\Gamma_{33\rightarrow 11}}\,
	\frac{\Gamma_{33\rightarrow 11}}{\Gamma_{22\rightarrow 00}} L^6
	\label{35}
\end{eqnarray}
where we have inserted the decay coefficients $\Gamma_{ii\rightarrow
jj}$
from level $i$ to level $j$ to be able to apply eq. (\ref{8}) which
then yields
\begin{equation}
	p_{em}^{(3)} = \frac{16\pi^2\epsilon^2\Gamma_{22}\Gamma_{33} }
	{\Delta_3^2 p_{em}^{(2)} } \left(\frac{E_{12}}{E_{02}}\right)^2
	\left(\frac{\omega_{02}}{\omega_{31}}\right)^3
	\frac{\Gamma_{33\rightarrow 11}}{\Gamma_{22\rightarrow 00}} L^6\;\;
,
	\label{36}
\end{equation}
where $E_{i2}$ is the electric field of the laser on the
$i\leftrightarrow 2$
transition. We now define the constant
\begin{equation}
	\alpha = \frac{\Gamma_{33}}{\Gamma_{22}}
	\frac{\Gamma_{22\rightarrow 00}}{\Gamma_{33\rightarrow 11}}
	\left(\frac{E_{02}}{E_{12}}\right)^2
	\label{37}
\end{equation}
and resolve eq. (\ref{36}) for $L$ to obtain
\begin{equation}
	L = \left\{
	\frac{\Delta_3^2 p_{em}^{(2)}p_{em}^{(3)}\alpha}
	{16\pi^2\epsilon^2\Gamma_{33}^2}
	\left(\frac{\omega_{13}}{\omega_{02}}\right)^3 \right\}^{1/6}\;\; .
	\label{38}
\end{equation}
Note that $E_{02}\ll E_{12}$ as argued below eq. (\ref{34}) and that
the
ratio of the decay coefficients in $\alpha$ will be of the order of
$1$.
Therefore $\alpha$ is much smaller than $1$. In fact, because of
$\Omega_{02}=\eta\Omega_{12}/\sqrt{5L}$, is a good approximation to
assume
$\alpha=\beta\eta^2/5L$ (with $\beta$ defined below eq. (\ref{39}))
which then yields
\begin{equation}
	L = \left\{
	\frac{\Delta_3^2 p_{em}^{(2)}p_{em}^{(3)}\eta^2\alpha}
	{80\pi^2\epsilon^2\Gamma_{33}^2}
	\left(\frac{\omega_{13}}{\omega_{02}}\right)^3 \right\}^{1/7}\;\; .
	\label{38a}
\end{equation}
Following very similar lines we obtain for case b)
\begin{equation}
	L = \left\{
	\frac{\Delta_3^2 p_{em}^{(2)}p_{em}^{(3)}\beta}
	{16\pi^2\epsilon^2\Gamma_{33}^2}
	\left(\frac{\omega_{13}}{\omega_{02}}\right)^3 \right\}^{1/6}\;\; .
	\label{39}
\end{equation}
where now $\beta=\Gamma_{33}\Gamma_{22\rightarrow 00}/
\Gamma_{22}\Gamma_{33\rightarrow 00}$ is of the order of $1$. Both
limits
eqs. (\ref{38}-\ref{39}) are analogous in their form to the estimates
eqs. (\ref{16}) and (\ref{17}) if we neglect in eqs. (\ref{16} -
\ref{17}) the
contribution from the upper level $1$ coupling to level $3$.

In table 2 we present some values for $L$ calculated from eqs.
(\ref{38a})
and (\ref{39}) for the same ions as in table 1. Again we assume
that $p_{em}^{(2)}=p_{em}^{(3)}=1$ (an optimistic choice) and observe
that the resulting limits for the numbers that can be factorized on a
quantum computer are very small, especially if $\eta=0.01$.
Nevertheless
one could think that transitions in $Ba^+$ and $Yb^+$ are promising.
However, two effects appear which lead
to further limitations. One is the fact that eqs. (\ref{38a}) and
(\ref{39})
are not necessarily more restrictive than the equivalent of eq.
(\ref{27})
for level $3$ as we have assumed eq. (\ref{32}) which does not need
to be
true. In fact simply applying eq. (\ref{27}) where $\Gamma_{22}$ is
replaced
by $\Gamma_{33}$ yields for $Hg^+$ $L=11$, for $Ba^+$ $L=20.5$ and
for $Yb^+$
$L=17.3$ which are somewhat smaller than the values for those ions in
table 2. Secondly and more importantly the promising results for
$Ba^+$
and $Yb^+$ are illusionary because a very stable transition is also
very
slow. From eq. (\ref{32a}) we can easily calculate that the
factorization
time using $Ba^+$ and assuming $\eta=1$ is
\begin{equation}
	T = 0.013 L^3
	\label{39a}
\end{equation}
which gives for $L=10$ a value of $T=13s$ during which no decohering
event
of any kind is allowed to happen. This is a very long time and it
will be difficult to isolate the quantum computer from the
environment. For $Yb^+$
the numbers are even more devastating. We obtain
\begin{equation}
	T= 51186 L^3
	\label{39b}
\end{equation}
which indicates for $L=4$ that $T=3.2\cdot 10^6 \approx 38 days$, a
completely
unrealistic number. This again shows very clearly that the solution
to the problem
of decoherence cannot be to employ extremely stable transitions
for qubits but to find methods that enable the quantum computer to
cope with
a certain level of decoherence, ie we need a form of quantum error
correction.
In principle it is indeed possible to implement quantum error
correction in
quantum computers but it remains the question inasmuch these methods
will
in fact improve the prospects of quantum computation. In the next
section we
will address this problem and derive limitations to quantum factoring
including the use of quantum error correction methods.
\section{Bounds on L including quantum error correction}
In the preceding section we derived intensity independent upper
bounds
for the numbers that can be factorized on a quantum computer. The
results
that we obtained are not very promising as they rule out the
possibility
to factorize large numbers (numbers with $L\le 400$ can be factorized
on
a classical computer) on a quantum computer. However,
in these estimates we have yet to include the possibility of quantum
error
correction
\cite{Shor2,Steane1,Steane2,Calderbank1,Ekert2,Knill1,Cirac2}
and fault-tolerant quantum computation
\cite{Shor3,DiVincenzo1,Plenio3}. These methods have the ability to
change
drastically the probability that the calculation becomes corrupted by
an
emission. For example, consider a quantum error
correction code that can correct a single general error. If the
probability
to suffer an error is $p\ll 1$ then the probability to suffer two
errors,
an event that cannot be corrected by the code, is $p^2$. Therefore,
we have
reduced the probability for a fatal error from $p$ to $p^2$ and it
is then possible to perform quadratically as many computational steps
as
possible without quantum error correction. Higher order correcting
codes
promise even higher gains in possible computational steps. However,
this simple
argument neglects two effects that counteract the expected benefits
of
quantum error correction codes. One reason is the quite obvious fact
that
the implementation of quantum error correction and fault-tolerant
quantum
computation requires substantial overheads in additional qubits for
encoding
the qubits and additional quantum gate operations to implement
quantum
logic operations on the encoded quantum bits. This overhead can be
quite
substantial even for simple quantum error correction codes.

The other effect is less obvious but can have an even stronger impact
on
the practical efficiency of quantum error correction codes. The
problem is
that quantum error correction codes loose their ability to correct
for
decohering events if population leaves the qubit as a result of the
decohering event. Therefore we have to hope that all population will
remain within the system representing the qubit. However, as we have
seen in
the analysis of section II this will not be the case for a realistic
implementation of a qubit. The reason is simply that the two-level
approximation is never perfectly satisfied. If we excite an
extraneous
level then an emission from this level does not necessarily bring the
population back into the qubit. An error of this kind cannot be dealt
with
by a quantum error correction code. One would have to find ways to
repump
the lost population back into the qubit. This, however, is quite
difficult
if not impossible because of the sheer multitude of possible levels
where
the spontaneous emission may lead the population to.

In this section we will first consider the case of quantum
computation
using two-level systems to represent the qubits. Subsequently we will
then consider the representation of qubits in Zeeman sublevels using
far
detuned Raman pulses for their manipulation.

Before we start with the discussion of the two-level system case, we
first
define two constants, $q$ and $c$, describing the necessary overheads
in
the number of qubits and the number of additional quantum gate
operations.
\begin{itemize}
\item The number of required qubits rises by a factor of $q$ as all
qubits
will have to be encoded and will be kept encoded during the
calculation.
Repeated decoding and encoding would raise the number of operations
substantially and lead to unnecessary errors.
\item The number of operations required to implement one quantum
logical operation on encoded qubits rises by a factor of $c$ because
we
have to perform more than one quantum gate to implement one quantum
logic
operation on the encoded qubits. in addition we also have to perform
quantum error corrections periodically.
\end{itemize}
The value of the factors $q$ and $c$ depends heavily on the quantum
error
correction code that is used. If one wants to protect a single qubit
against one general error, then it is known that $q=5$ is the minimal
possible value \cite{Knill1}. The number $c$ for this code is
uncertain
in general for the most optimal codes but is at least $c=5$ for a
CNOT
gate \cite{Shor3} and probably much more for other gates such as a
Toffoli
gate.
\subsection{Qubits as two-level systems}
For the following analysis we assume an error correction code capable
of correcting one error. A similar calculation will then yield the
result
for codes that can correct $k-1$ errors and we simply state these
results.
First we need to find the time required to perform a CNOT operation
on encoded quantum bits. The computer contains $q$ times more qubits,
ie. $5qL$ qubits, and $c$ times more operations are required.
Therefore
\begin{equation}
	\tau_{el} = \frac{4\pi\sqrt{5qL}}{\eta\Omega_{01}}\, c\; .
	\label{40}
\end{equation}
The spontaneous lifetime of the quantum computer is then
\begin{equation}
	\tau_{sp} = \frac{1}{5Lq\Gamma_{11}}\;\; .
	\label{41}
\end{equation}
As we now have the ability to correct for errors during the
calculation,
we no longer need to require that there is practically no spontaneous
emission  during the whole calculation. Instead we ask for the
probability
$p_N$ to have one error during $N$ logical operations of our
algorithm to
be small compared to $1$. We find
\begin{equation}
	p_N = \frac{4\pi\sqrt{5qL}}{\eta\Omega_{01}}\, 5cqL\Gamma_{11} N
	\;\; .
	\label{42}
\end{equation}
If we perform a fault-tolerant error correction after these N
operations
then this correction will fail with probability $p_N^2$, the
probability
to have suffered two errors, because the code is not designed to
correct
for two errors. Therefore the probability that the whole computation
fails
is given by
\begin{equation}
	p_{fail} = p_N^2 \frac{\epsilon L^3}{N}\;\; .
	\label{43}
\end{equation}
This immediately implies that $N=1$ is the optimal choice unless we
take
the $N$ dependence of $c$ into account which is however quite
difficult
as $c$ is not known very well. For the following we will assume $N=1$
and
we obtain
\begin{equation}
	L = \left(\frac{\eta^2 p_{fail}}{2000\pi^2 q^3 c^2 \epsilon}
	\left(\frac{\Omega_{01}}{\Gamma_{11}}\right)^2 \right)^{1/6}
	\;\; .
	\label{44}
\end{equation}
{}From eq. (\ref{44}) we easily deduce the total computation time
\begin{equation}
	T = \frac{400\pi^2 q^2 c^2 \epsilon^{3/2}}
	{\eta^2 p_{fail}^{1/2} } \frac{\Gamma_{11}}{\Omega_{01}^2}
	L^{6.5} \;\; .
	\label{45}
\end{equation}
Again eqs. (\ref{44}) and (\ref{45}) are dependent on the laser power
driving the system. To eliminate this intensity dependence
we now have to take other levels into account as we did in
section II. Again we have to distinguish the two cases a) (qubit
transition
is quadrupole allowed) and b) (qubit transition is octupole allowed)
depending on the nature of the qubit transition. We first treat case
a)
and then state the result for case b).

To keep the resulting expressions more transparent we neglect the
contribution
of the upper qubit level $1$ to the violation of the two-level
approximation. The generalization to this case is easy.We then find
for the
population in the extraneous level
$2$
\begin{equation}
	\rho_{22} = \frac{\Omega_{02}^2}{8\Delta_2^2}\;\; .
	\label{46}
\end{equation}
Now we want to know the probability $p_{out}$ that an emission from
level $2$ which leads to population {\em outside} the qubit
transition
occurs during the whole computation. It is very important to note
that
it is this probability $p_{out}$ that has to be much smaller than
unity
because even a single emission leading out of the qubit cannot be
corrected by the quantum error correction code. We find
\begin{equation}
	p_{out} =
	2\Gamma_{22}^{out} \frac{\Gamma_{22\rightarrow 00}\Omega_{02}^2}
	{8\Delta_2^2 \Gamma_{22\rightarrow 00}} \,
	\frac{400\pi^2 q^2 c^2 \epsilon^{3/2}}
	{\eta^2 p_{fail}^{1/2} } \frac{\Gamma_{11}}{\Omega_{01}^2}
	L^{6.5} \;\; .
	\label{47}
\end{equation}
This can be solved for $L$ and with eq. (\ref{8}) we obtain
\begin{equation}
	L = \left\{ \frac{\eta^2\Delta_2^2 \, p_{fail}^{1/2} \, p_{out}}
	{100\pi^2 q^2 c^2 \epsilon^{3/2} \Gamma_{22\rightarrow 00}
	\Gamma_{22}^{out} }
	\left(\frac{\omega_{02}}{\omega_{01}} \right)^3
	\right\}^{\frac{2}{13}} \;\; .
	\label{48}
\end{equation}
With an analogous calculation we obtain for case b)
\begin{equation}
	L = \left\{ \frac{\Delta_2^2 \, p_{fail}^{1/2} \, p_{out}}
	{20\pi^2 q^2 c^2 \epsilon^{3/2} \Gamma_{22\rightarrow 00}
	\Gamma_{22}^{out} }
	\frac{\omega_{02}}{\omega_{01}}
	\right\}^{\frac{2}{11}} \;\; .
	\label{49}
\end{equation}
The estimates eqs. (\ref{48}) and (\ref{49}) are valid for a quantum
error
correction code that is able to correct for one error. One can quite
easily extend these results to quantum error correction codes that
are
able to correct for $k-1$ errors. We state only the main results for
the cases a) and b) as they can be derived easily along the lines
that
lead us to eqs. (\ref{48}) and (\ref{49}). For case a) we obtain the
computation time
\begin{equation}
	T = \frac{400\pi^2 c^2 q^2 \epsilon}{\eta^2}
	\left(\frac{\epsilon}{p_{fail}} \right)^{\frac{1}{k}}
	\frac{\Gamma_{11}}{\Omega_{01}^2} L^{5+\frac{3}{k}} \;\; .
	\label{50}
\end{equation}
Now we take into account other levels outside the qubit transition.
A spontaneous emission that leads from these levels out of the
qubit system cannot be corrected for by the quantum error correction
code. Therefore we obtain
\begin{equation}
	L = \left\{ \frac{\Delta_2^2\eta^2 p_{out}}{100\pi^2 c^2 q^2
	\epsilon \, \Gamma_{22}^{out}\Gamma_{22\rightarrow 00} }
	\left( \frac{\omega_{02}}{\omega_{01}} \right)^3
	\left( \frac{p_{fail}}{\epsilon} \right)^{\frac{1}{k}}
	\right\}^{\frac{k}{5 k + 3}} \;\; .
	\label{51}
\end{equation}
For case b) we obtain the result
\begin{equation}
	L = \left\{ \frac{\Delta_3^2 p_{out}}{20\pi^2 c^2 q^2 \epsilon \,
	\Gamma_{22}^{out}\Gamma_{22\rightarrow 00} }
	\frac{\omega_{02}}{\omega_{01}}
	\left( \frac{p_{fail}}{\epsilon} \right)^{\frac{1}{k}}
	\right\}^{\frac{k}{4 k + 3}} \;\; .
	\label{52}
\end{equation}
Again we can now discuss numerical values for eqs. (\ref{48}) and
(\ref{49}) obtained from the atomic data of real ions. In table 3 we
give the results for eqs. (\ref{48}) and (\ref{49}). we see that
although the estimates have improved they still restrict
factorization
to very small numbers especially for a Lamb-Dicke parameter
$\eta=0.01$
where we obtain upper limits that restrict the numbers that are
possible
to factorize to trivial sizes. Again $Yb^+$ seems to be very
promising
but again we should note that it is very hard to actually drive the
this
transition sufficiently quickly so that the computation times quickly
reach astronomical values. For $\Omega_{01}^2/\Gamma_{11}=10^{16}$ we
obtain for $L=4$ a value of $T=1400s$, a value which is extremely
high.
For $Ba^+$ we obtain for $L=4$, $\Omega_{01}^2/\Gamma_{11}=10^{16}$
and
$\eta=1$ the value $T=0.84s$.
\subsection{Qubits in Zeeman sublevels}
Now we perform the same analysis including quantum error correction
codes for the case where the qubits is stored in Zeeman sublevels.
Of course the analysis here runs along similar lines as the one for
the two-level system and that of section II. Therefore we only state
the final result for the cases a) and b). We present the results for
an error correcting code which is able to correct $k-1$ errors. We
only
treat the case where the extraneous levels give a small correction to
the time evolution of the system, ie we assume the regime of eq.
(\ref{31}).
For the computation time we obtain the lower limit
\begin{equation}
	T \gg \frac{8\pi\epsilon L^3 \Gamma_{33\rightarrow 00}}
	{\Gamma_{22\rightarrow 00}\Delta_3} c
	\label{53}
\end{equation}
for both cases a) and b). For case a) we obtain the estimate
\begin{equation}
	L = \left\{ \frac{\alpha \, p_{em}^{(3)} \, \Delta_{3}^2 }
	{16\pi^2 c^2 \epsilon \Gamma_{33}^{out}\Gamma_{33\rightarrow 11}}
	\left( \frac{\omega_{13}}{\omega_{02}} \right)^3
	\left( \frac{ p_{fail} }{ \epsilon } \right)^{\frac{1}{k}}
	\right\}^{ \frac{k}{3k+3} } \;\; ,
	\label{54}
\end{equation}
where
\begin{equation}
	\alpha = \frac{\Gamma_{22\rightarrow 00}}{\Gamma_{22}}
	\left( \frac{E_{02}}{E_{12}} \right)^2 \ll 1 \;\; .
	\label{55}
\end{equation}
$E_{i2}$ denotes the electric field strength of the laser on the
$i \leftrightarrow 2$ transition. Note that again we have to take
into
account that $\alpha$ contains a hidden dependence on $\eta,L$ and
$q$.
This dependence is due to the fact that we have to satisfy the
condition
\begin{equation}
	\Omega_{02} = \frac{\eta\Omega_{12}}{\sqrt{5Lq}}\;\; .
	\label{55a}
\end{equation}
Therefore $\alpha\cong\beta\eta^2/5Lq$ ($\beta$ is defined in
(\ref{57}))
and we obtain
\begin{equation}
	L = \left\{ \frac{\beta \, p_{em}^{(3)} \, \Delta_{3}^2 }
	{80\pi^2 c^2 \epsilon \Gamma_{33}^{out}\Gamma_{33\rightarrow 11}}
	\left( \frac{\omega_{13}}{\omega_{02}} \right)^3
	\left( \frac{ p_{fail} }{ \epsilon } \right)^{\frac{1}{k}}
	\right\}^{ \frac{k}{4k+3} } \;\; .
	\label{58}
\end{equation}
For case b) we obtain a very
similar result. We find
\begin{equation}
	L = \left\{ \frac{\beta \, p_{em}^{(3)} \, \Delta_{3}^2 }
	{ 32\pi^2 c^2 \epsilon \,\Gamma_{33}^{out}\Gamma_{33\rightarrow 11}
}
	\left( \frac{\omega_{03}}{\omega_{02}} \right)^3
	\left( \frac{ p_{fail} }{ \epsilon } \right)^{\frac{1}{k}}
	\right\}^{\frac{k}{3k+3}} \;\; ,
	\label{56}
\end{equation}
where
\begin{equation}
	\beta = \frac{ \Gamma_{22\rightarrow 00} }{ \Gamma_{22} }
	\approx \frac{1}{2} \;\; .
	\label{57}
\end{equation}
 Assuming that $p_{em}^{(3)}\beta=1$ we then obtain the values given
in
table 4 which again seem to be promising if we assume $\eta=1$. For
the more realistic values of $\eta=0.01$, however, the estimate
reduces substantially. In the discussion of table 2 we already
pointed out
that high values for the estimates on $L$ alone do not suffice to
allow
a practical implementation of quantum factoring. Important also is
that
the computation time be sufficiently short. This restriction again
excludes
$Yb^+$ immediately because for
$\Omega_{02}^2/\Gamma_{22\rightarrow 00}=10^{16},\eta=1$ and $L=4$ we
obtain $T=189$days. It also makes $Ba^+$ a very unlikely candidate
for
successful factorization as for $L=10$ we obtain with the same
parameters
as for $Yb^+$ the value $T=65s$.
\section{Conclusions}
In this paper we have discussed the constraints imposed by
spontaneous
emission onto factorization of big numbers using a quantum computer.
In
section II we have investigated the possibility of factorization
without the use of quantum error correction. We were able to derive
limits
to the bitsize of the numbers that can be factorized on a quantum
computer.
These limits are {\em independent} of the intensity of the laser that
is
used to implement quantum gates. This intensity independence was
achieved
by taking into account the failure of the two-level approximation
even
for modest laser powers. The result of these limits is that without
the use of quantum error correction the factorization even of small
numbers will not be possible.
Therefore we then investigated in section III whether the
implementation of quantum error correction can improve the prospects
of factorization on a quantum computer. It turned out that quantum
error
correction in realistic atomic systems is much less effective than in
the ideal case of closed two-level systems because the failure of the
two-level approximation can lead to a leakage of population out of
the
system which can not be corrected by a quantum error correction code.
Therefore even high order quantum error correction codes are
susceptible
to single errors that lead to loss of population out of the system
which
corrupts their performance. Nevertheless the upper limits in this
case
are higher than without quantum error correction. However, the total
computation time using quantum error correction, and even in the case
without quantum error correction, tends to be very large because a
very stable transition cannot be driven very fast. We find lower
bounds
for the computation times and conclude that even for moderate numbers
it will be difficult to isolate the system from the environment such
that
no decohering event, other than a spontaneous emission, can take
place.
At this point other decoherence effects have to be taken into account
(see eg. \cite{Garg1,Hughes1,James1}). However, we think that even
the sole inclusion of spontaneous emission as a decohering effect
already
shows that the practical application of quantum computers in
factorization
of big numbers will be highly unlikely within the present models of
quantum computation. The considerations of this paper also apply to
other possible algorithms for quantum computers if they
also require $O(L^3)$ operations.

Although the conclusions of this paper are rather pessimistic with
regard to the practical application of quantum computers for actual
computations we would like to stress that there are applications such
as
for example in precision spectroscopy \cite{Wineland1} which require
much less operations, $O(L)$, and for which therefore the above
presented
estimates will be much more optimistic. We believe that in this area
interesting applications of the concept of quantum computing will be
found.
However, unrealistic hopes that have been raised in the past about
the
practicality of quantum computers have to be moderated.

\section{Acknowledgements}
We would like to thank C. Monroe and J.I. Cirac for alerting our
attention to the method of employing Raman transitions to drive
qubits.
We acknowledge discussions with D.F.V. James and R.J. Hughes and
thank
them for providing us with the experimental data for the ions.
This work was supported by a European Community Network, the UK
Engineering and Physical Sciences
Research Council and by a Feodor-Lynen grant of the Alexander von
Humboldt Foundation\\[1.cm]

\newpage

\begin{center}
	{\bf FIGURE CAPTIONS}\\[.5cm]
\end{center}

\begin{description}
\begin{minipage}[t]{1.cm}\item{Fig. 1 :}\end{minipage}\hfill
\begin{minipage}[t]{14.cm}
A two-level system representing a qubit. The levels $0$ and $1$
represent
the two possible logical values. Unlike in classical computing
coherent
superpositions between the logical values $0$ and $1$ are possible.
The upper level may decay spontaneously
to the ground state with a decay rate $2\Gamma_{11}$ and the system
can be driven by a laser with Rabi frequency $\Omega_{01}$.
\end{minipage}\\[.5cm]
\begin{minipage}[t]{1.cm}\item{Fig. 2 :}\end{minipage}\hfill
\begin{minipage}[t]{14.cm}
The two lower levels $0$ and $1$ of the $\Lambda$ system,
are Zeeman sublevels of a stable ground state and represent the
logical
values $0$ and $1$ of the qubit. They are coupled via Raman pulses
that
are strongly detuned from the intermediate level $2$. The two-photon
detuning is assumed to be zero while the one photon detuning
$\Delta_2$
is much larger than the associated Rabi frequencies on the
$i\leftrightarrow 2$ transition is $\Omega_{i2}$.
\end{minipage}\\[.5cm]
\begin{minipage}[t]{1.cm}\item{Fig. 3 :}\end{minipage}\hfill
\begin{minipage}[t]{14.cm}
Schematic picture of the excitation of several ions in a linear
ion trap. The translational degrees of freedom of the ions are
assumed to be cooled to their respective ground states. To
implement quantum gates, standing wave fields interact with the
ions and thereby change the inner state of the ions as well as
the state of the center-of-mass mode (which leads to entanglement).
\end{minipage}\\[.5cm]
\begin{minipage}[t]{1.cm}\item{Fig. 4 :}\end{minipage}\hfill
\begin{minipage}[t]{14.cm}
Schematic level scheme envisaged for quantum computation, in which
the $0\leftrightarrow 1$ transition represents the qubit. It is
driven by
a laser of Rabi frequency $\Omega_{01}$. Level $1$ has a spontaneous
decay
rate $2\Gamma_{11}$ which is small. The laser which is resonant with
the
$0\leftrightarrow 1$ transition inevitably couples level $0$ also to
other non-resonant levels as for example level $2$. The Rabi
frequency on
that transition is then $\Omega_{02}$ and the decay rate
is $2\Gamma_{22}$ is usually much larger than $\Gamma_{11}$. The
effective
Rabi frequency on the $0\leftrightarrow 2$ transition is very small
as the laser is detuned by $\Delta_{02}\gg \Omega_{02}$.
\end{minipage}\\[.5cm]
\begin{minipage}[t]{1.cm}\item{Fig. 5 :}\end{minipage}\hfill
\begin{minipage}[t]{14.cm}
Results of a discrete Fourier transform (DFT) of a function
$f(n)=\delta_{8,(n mod 10)}$ with $n=0,1,\ldots,31$. The solid line
is the result for a quantum computer with stable qubits and
represents
the correct result. The dashed line shows the result of the same
computation using a quantum computer with unstable qubits, one of
which
has suffered a spontaneous emission during the calculation. The
results
clearly differ and show the impact of a single spontaneous emission
on a
quantum computation. For the parameters chosen on average the quantum
computer will suffer one emission per DFT, ie $\tau_{sp}=T$ in this
case.
\end{minipage}\\[.5cm]
\begin{minipage}[t]{1.cm}\item{Fig. 6 :}\end{minipage}\hfill
\begin{minipage}[t]{14.cm}
The same quantum computation as in Fig. 5. The solid line again
represents
the result using a quantum computer with stable qubits, while the
dashed
line shows the result using a quantum computer with unstable qubits.
This
time, however, the unstable quantum computer does not suffer an
emission
during the whole calculation. Again the results differ illustrating
the
impact of the conditional time evolution between spontaneous
emissions.
\end{minipage}\\[.5cm]
\begin{minipage}[t]{1.cm}\item{Table 1 :}\end{minipage}\hfill
\begin{minipage}[t]{14.cm}
For several possible systems the upper limit on the bitsize
$L$ of the number $N$ that can be factorized on a quantum computer
is calculated. A qubit is stored in a metastable optical transition.
The atomic levels which are abbreviated in Fig. \protect\ref{Fig1} by
$0,1$
and $2$ are given. The atomic data are inserted
into eqs. (\ref{16}) and (\ref{17}) and the result is given in the
last
row of the table. The atomic data are inserted into eqs. (\ref{16})
and
(\ref{17}) and the result is given in the last row of the table. The
atomic
data are derived from (A) = \cite{Bashkin1}, (B) = \cite{Bergquist1},
(C) = \cite{Sauter1,Sauter2}, (D) = calculated using\cite{Bell1} and
\cite{Gill1}, (E) = \cite{Andersen1},
(F) = \cite{Fawcett1},(G) = \cite{Gosselin1}, (H) = \cite{Eriksen1},
(I) = \cite{Gallagher1}, (J) = \cite{Wiese1}.
\end{minipage}\\[.5cm]
\begin{minipage}[t]{1.cm}\item{Table 2 :}\end{minipage}\hfill
\begin{minipage}[t]{14.cm}
For several possible systems the upper limit on the bitsize
$L$ of the number $N$ that can be factorized on a quantum computer
is calculated. A qubit is stored in two Zeeman sublevels of a stable
ground state. The atomic levels which are abbreviated in Fig.
\protect\ref{Fig7} by
$0,1,2$ and $3$ are given. The atomic data are inserted
into eq. (\ref{21}) and the result is given in the last rows of the
table.
\end{minipage}\\[.5cm]
\begin{minipage}[t]{1.cm}\item{Fig. 7 :}\end{minipage}\hfill
\begin{minipage}[t]{14.cm}
The $\Lambda$ system as in Fig. 2 but now the coupling to an
extraneous levels $3$ is taken into account. The lasers performing
the Raman pulses also couples the two lower states to level $3$
which may subsequently decay spontaneously.
\end{minipage}\\[.5cm]
\begin{minipage}[t]{1.cm}\item{Table 3 :}\end{minipage}\hfill
\begin{minipage}[t]{14.cm}
For several possible systems the upper limit on the bitsize
$L$ of the number $N$ that can be factorized on a quantum computer
using quantum error correction is calculated. A qubit is stored in a
metastable optical transition.The atomic levels which
are abbreviated in Fig. \protect\ref{Fig7} by
$0,1$ and $2$ are given. The atomic data  are inserted
into eq. (\ref{21}) and the result is given in the last rows of the
table.
\end{minipage}\\[.5cm]
\begin{minipage}[t]{1.cm}\item{Table 4 :}\end{minipage}\hfill
\begin{minipage}[t]{14.cm}
For several possible systems the upper limit on the bitsize
$L$ of the number $N$ that can be factorized on a quantum computer
is calculated. A qubit is stored in two Zeeman sublevels of a stable
ground state. The atomic levels which are abbreviated in Fig.
\protect\ref{Fig7} by
$0,1,2$ and $3$ are given. The atomic data are inserted
into eq. (\ref{21}) and the result is given in the last rows of the
table.
\end{minipage}
\end{description}

\newpage

\begin{figure}[hbt]

\setlength{\unitlength}{1.1mm}
\begin{picture}(100,75)
\thicklines
\put(40,0){\line(1,0){30}}
\put(40,0.3){\line(1,0){30}}
\put(40,30){\line(1,0){30}}
\put(40,30.3){\line(1,0){30}}
\put(55,3){\vector(0,1){26}}
\put(55,27){\vector(0,-1){26}}
\multiput(53,27.5)(0,-1.75){14}{\makebox(0,0)[c]{$\wr$}}
\multiput(52.8,27.5)(0,-1.75){14}{\makebox(0,0)[c]{$\wr$}}
\put(52.9,6){\vector(0,-1){5}}
\large
\put(47.5,15){\makebox(0,0)[c]{$\Gamma_{11}$}}
\put(60,15){\makebox(0,0)[c]{$\Omega_{01}$}}
\put(55,-5){\makebox(0,0)[c]{$0$}}
\put(55,34){\makebox(0,0)[c]{$1$}}
\normalsize
\end{picture}
\vspace*{2.cm}
\begin{center}
	Plenio and Knight FIGURE 1
\end{center}
\caption{\label{Fig1}}
\end{figure}

\newpage

\begin{figure}[hbt]
\setlength{\unitlength}{1.0mm}
\begin{picture}(50,40)
\thicklines
\put(10,-10){\line(1,0){30}}
\put(10,-9.7){\line(1,0){30}}
\put(5,-10){\makebox(5,5)[bl]{0}}
\put(25,32){\line(1,0){30}}
\put(25,32.3){\line(1,0){30}}
\put(20,32){\makebox(5,5)[bl]{2}}
\put(40,-6){\line(1,0){30}}
\put(40,-5.7){\line(1,0){30}}
\put(75,-6){\makebox(5,5)[bl]{1}}
\put(24.8,-7){\vector(1,3){10}}
\put(34,21){\vector(-1,-3){10}}
\put(20,7){\makebox(5,5)[bl]{$\Omega_{02}$}}
\put(57,-4){\vector(-1,3){9}}
\put(48.2,22){\vector(1,-3){9}}
\put(56,7){\makebox(5,5)[bl]{$\Omega_{12}$}}
\multiput(25,24)(5.4,0){6}{\line(1,0){3}}
\multiput(25,24.3)(5.4,0){6}{\line(1,0){3}}
\Huge
\put(60,23.5){\makebox(5,5)[bl]{$\left.\right\}$}}
\normalsize
\put(67,26){\makebox(5,5)[bl]{$\Delta_2$}}
\end{picture}
\vspace*{2.cm}
\begin{center}
	Plenio and Knight FIGURE 2
\end{center}
\caption{\label{Fig2}}
\end{figure}

\newpage

\begin{figure}[hbt]
\setlength{\unitlength}{1.0mm}
\begin{picture}(100,50)
\thicklines
\put(0,0){\line(1,0){10}}
\put(0,0){\line(0,1){5}}
\put(0,5){\line(1,0){10}}
\put(10,2.5){\oval(10,5)[r]}
\multiput(25,2.5)(15,0){8}{\circle*{5}}
\multiput(25,-3)(15,0){8}{\vector(1,0){5}}
\multiput(25,-3)(15,0){8}{\vector(-1,0){5}}
\multiput(23,20)(15,0){8}{\line(0,1){10}}
\multiput(27,20)(15,0){8}{\line(0,1){10}}
\multiput(23,30)(15,0){8}{\line(1,0){4}}
\multiput(23,20)(15,0){8}{\line(1,0){1}}
\multiput(26,20)(15,0){8}{\line(1,0){1}}
\multiput(39.5,29)(.2,0){5}{\line(0,-1){22}}
\multiput(99.5,29)(.2,0){5}{\line(0,-1){22}}
\large
\put(49,35){\makebox(5,5)[bl]{Standing wave laser fields}}
\normalsize
\put(145,2.5){\oval(10,5)[l]}
\put(145,0){\line(1,0){10}}
\put(155,0){\line(0,1){5}}
\put(155,5){\line(-1,0){10}}
\end{picture}
\vspace*{2.cm}
\begin{center}
	Plenio and Knight FIGURE 3
\end{center}
\caption{\label{Fig3}}
\end{figure}

\newpage

\begin{figure}[hbt]

\setlength{\unitlength}{1.0mm}
\begin{picture}(100,75)
\thicklines
\put(10,0){\line(1,0){60}}
\put(10,0.3){\line(1,0){60}}
\put(40,30){\line(1,0){30}}
\put(40,30.3){\line(1,0){30}}
\put(10,60){\line(1,0){30}}
\put(10,60.3){\line(1,0){30}}
\put(55,3){\vector(0,1){26}}
\put(55,27){\vector(0,-1){26}}
\put(26,3){\vector(0,1){56}}
\put(26,57){\vector(0,-1){56}}
\multiput(23,57.5)(0,-1.75){32}{\makebox(0,0)[c]{$\wr$}}
\multiput(22.8,57.5)(0,-1.75){32}{\makebox(0,0)[c]{$\wr$}}
\put(22.9,6){\vector(0,-1){5}}
\multiput(53,27.5)(0,-1.75){14}{\makebox(0,0)[c]{$\wr$}}
\multiput(52.8,27.5)(0,-1.75){14}{\makebox(0,0)[c]{$\wr$}}
\put(52.9,6){\vector(0,-1){5}}
\large
\put(47.5,15){\makebox(0,0)[c]{$\Gamma_{11}$}}
\put(17.5,30){\makebox(0,0)[c]{$\Gamma_{22}$}}
\put(31,30){\makebox(0,0)[c]{$\Omega_{02}$}}
\put(60,15){\makebox(0,0)[c]{$\Omega_{01}$}}
\put(-.5,60){\makebox(0,0)[c]{$E=\hbar\omega_{02}$}}
\put(80,30){\makebox(0,0)[c]{$E=\hbar\omega_{01}$}}
\put(40,-5){\makebox(0,0)[c]{$0$}}
\put(25,64){\makebox(0,0)[c]{$2$}}
\put(55,34){\makebox(0,0)[c]{$1$}}
\normalsize
\end{picture}
\vspace*{2.cm}
\begin{center}
	Plenio and Knight FIGURE 4
\end{center}
\caption{\label{Fig4}}
\end{figure}

\newpage

\begin{figure}[hbt]
\epsfxsize14.cm
\centerline{\epsffile{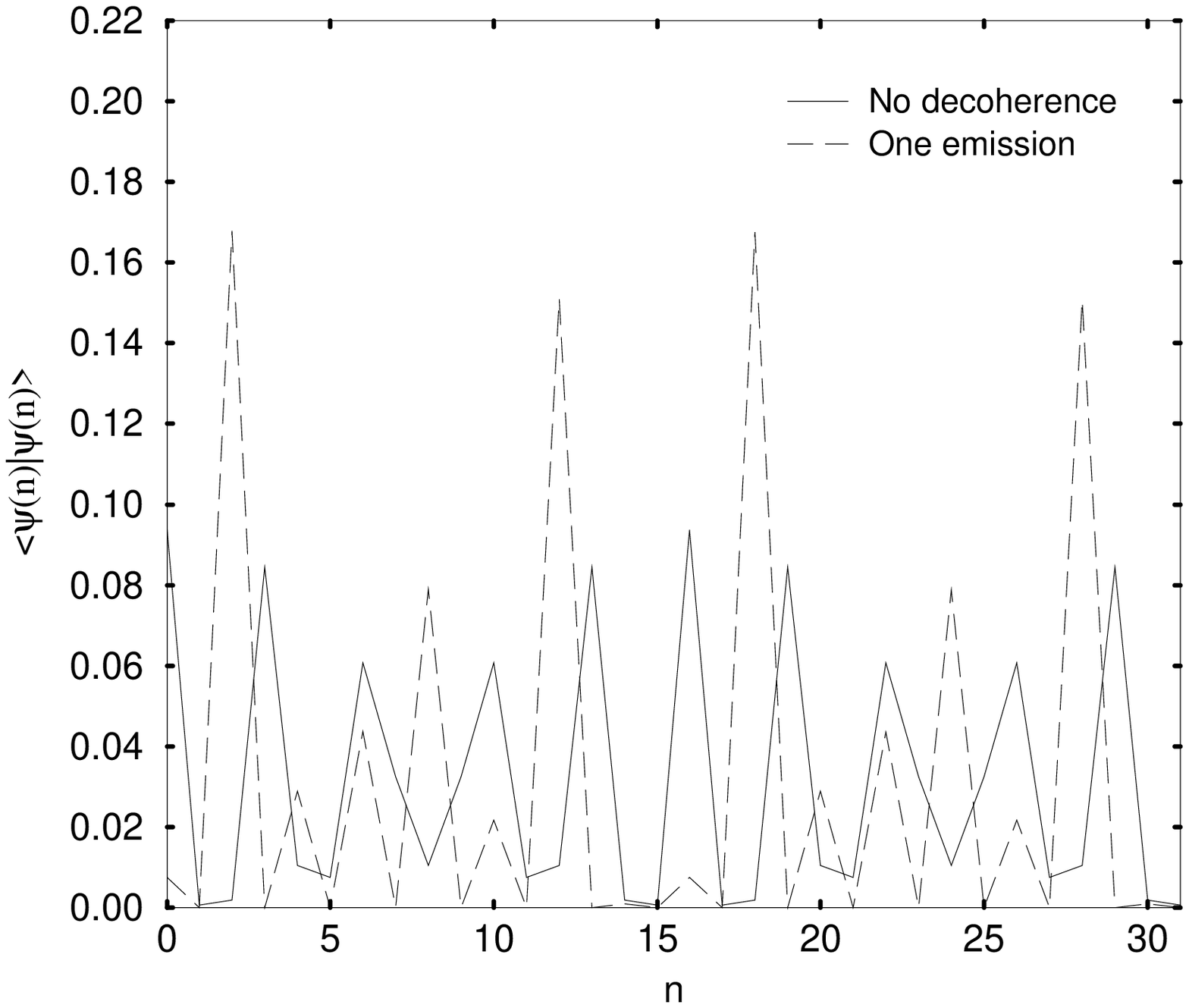}}
\caption{\label{Fig5}}
\end{figure}
\begin{center}
	Plenio and KnightFIGURE 5
\end{center}

\newpage

\begin{figure}[hbt]
\epsfxsize14.cm
\centerline{\epsffile{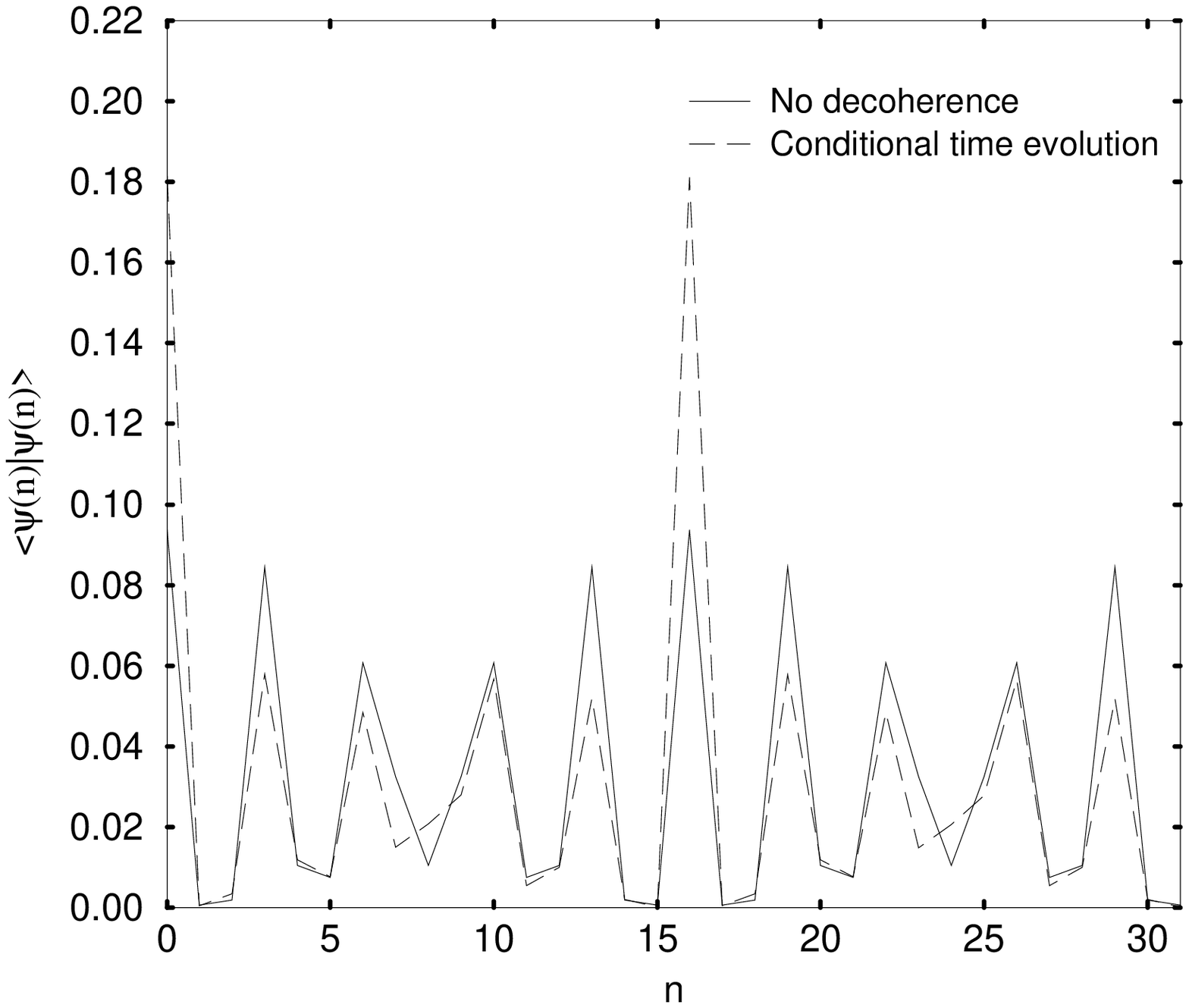}}
\caption{\label{Fig6}}
\end{figure}
\begin{center}
	Plenio and Knight FIGURE 6
\end{center}

\newpage

\begin{center}
\begin{tabular}{|c||c|c|c|c|}\hline
Ion                          &         $Ca^+$         &
$Hg^+$             &         $Ba^+$         &         $Yb^+$
      \\ \hline
level 0                      & $4s\, {}^2 S_{1/2}$    &
$5d^{10}6s^{2}\, {}^2 S_{1/2}$ &  $6s \,{}^2 S_{1/2}$   & $4f^{14}
6s^2\, {}^2 S_{1/2}$ \\ \hline
level 1                      & $3d\, {}^2 D_{5/2}$    &
$5d^{9}6s^{2}\, {}^2 D_{5/2}$  &  $5d \,{}^2 D_{5/2}$   & $4f^{13}
6s^2\, {}^2 F_{7/2}$ \\ \hline
level 2                      & $4s\, {}^2 P_{3/2}$    &
$5d^{10}6p^{2}\, {}^2 P_{1/2}$ &  $6s \,{}^2 P_{3/2}$   & $4f^{14}
6p^2\, {}^2 P_{3/2}$ \\ \hline\hline
$\;\omega_{01}\,[s^{-1}]\;$  & $2.61\cdot 10^{15}$ (A) &
                               $6.7\cdot 10^{15}$  (B) &
                               $1.07\cdot 10^{15}$ (C) &
			       $4.04\cdot 10^{15}$ (D)\\ \hline
$\;\omega_{02}\,[s^{-1}]\;$  & $4.76\cdot 10^{15}$ (A) &
		               $11.4\cdot 10^{15}$ (E) &
                               $4.14\cdot 10^{15}$ (C) &
                               $5.7\cdot 10^{15}$  (F)  \\
\hline\hline
$\;\Gamma_{22\rightarrow 00}\,[s^{-1}]\;$ & $67.5\cdot 10^6$ (G) &
                                            $5.26\cdot 10^8$ (H) &
                                            $58.8\cdot 10^6$ (I) &
                                            $60.2\cdot 10^6$ (F) \\
\hline
$\;\Gamma_{22\rightarrow 11}\,[s^{-1}]\;$ & $4.95\cdot 10^6$ (J) &
                                            $1.25\cdot 10^8$ (H) &
                                            $18.5\cdot 10^6$ (I) &
                                            $0.03$ (F)     \\
\hline\hline
$\;\;L(\eta=1)\;\;$          &      $6.9$             &
$4.9$              &  $14.2$                &  $14.3$
\\ \hline
$\;\;L(\eta=0.01)\;\;$       &      $2.2$             &
$1.6$              &  $4.5$                 &  $14.3$
\\ \hline
\end{tabular}
\end{center}
\vspace*{1.cm}
\begin{center}
	Plenio and Knight TABLE 1
\end{center}

\newpage

\begin{center}
\begin{tabular}{|c||c|c|c|c|}\hline
Ion                          &         $Ca^+$         &
$Hg^+$             &         $Ba^+$         &         $Yb^+$
      \\ \hline
level 0                      & $4s\, {}^2 S_{1/2}$    &
$5d^{10}6s^{2}\, {}^2 S_{1/2}$ &  $6s \,{}^2 S_{1/2}$   & $4f^{14}
6s^2\, {}^2 S_{1/2}$ \\ \hline
level 2                      & $3d\, {}^2 D_{5/2}$    &
$5d^{9}6s^{2}\, {}^2 D_{5/2}$  &  $5d \,{}^2 D_{5/2}$   & $4f^{13}
6s^2\, {}^2 F_{7/2}$ \\ \hline
level 3                      & $4s\, {}^2 P_{3/2}$    &
$5d^{10}6p^{2}\, {}^2 P_{1/2}$ &  $6s \,{}^2 P_{3/2}$   & $4f^{14}
6p^2\, {}^2 P_{3/2}$ \\ \hline\hline
$\;\omega_{02}\,[s^{-1}]\;$  & $2.61\cdot 10^{15}$    &
			       $6.7\cdot 10^{15}$     &
			       $1.07\cdot 10^{15}$    &
			       $4.04\cdot 10^{15}$   \\ \hline
$\;\omega_{13}\,[s^{-1}]\;$  & $4.76\cdot 10^{15}$   &
 			       $11.4\cdot 10^{15}$   &
	                       $4.14\cdot 10^{15}$   &
	                       $5.7\cdot 10^{15}$    \\ \hline\hline
$\;\Gamma_{33}\,[s^{-1}]\;$  & $73\cdot 10^6$ (J)   &
		               $6.51\cdot 10^8$ (H) &
                               $79.7\cdot 10^6$ (I) &
                               $60.2\cdot 10^6$	(F) \\ \hline\hline
$\;\;L(\eta=1)\;\;$          &      $14.0$            &
$4.2$              &  $24.4$                &  $26.0$
\\ \hline
$\;\;L(\eta=0.01)\;\;$       &      $4.0$             &
$1.4$              &  $6.4$                 &  $26.0$
\\ \hline
\end{tabular}
\end{center}
\vspace*{1.cm}
\begin{center}
	Plenio and Knight TABLE 2
\end{center}

\newpage

\begin{figure}[hbt]
\setlength{\unitlength}{1.0mm}
\begin{picture}(50,40)
\thicklines
\put(10,-10){\line(1,0){30}}
\put(10,-9.7){\line(1,0){30}}
\put(5,-10){\makebox(5,5)[bl]{0}}
\put(25,32){\line(1,0){30}}
\put(25,32.3){\line(1,0){30}}
\put(20,32){\makebox(5,5)[bl]{2}}
\put(40,-6){\line(1,0){30}}
\put(40,-5.7){\line(1,0){30}}
\put(75,-6){\makebox(5,5)[bl]{1}}
\put(24.8,-7){\vector(1,3){10}}
\put(34,21){\vector(-1,-3){10}}
\put(20,7){\makebox(5,5)[bl]{$\Omega_{02}$}}
\put(57,-4){\vector(-1,3){9}}
\put(48.2,22){\vector(1,-3){9}}
\put(56,7){\makebox(5,5)[bl]{$\Omega_{12}$}}
\multiput(25,24)(5.4,0){6}{\line(1,0){3}}
\multiput(25,24.3)(5.4,0){6}{\line(1,0){3}}
\Huge
\put(60,23.5){\makebox(5,5)[bl]{$\left.\right\}$}}
\normalsize
\put(67,26){\makebox(5,5)[bl]{$\Delta_2$}}
\put(25,50){\line(1,0){30}}
\put(25,50.3){\line(1,0){30}}
\put(20,49){\makebox(5,5)[bl]{$3$}}
\end{picture}
\vspace*{2.cm}
\begin{center}
	Plenio and Knight FIGURE 7
\end{center}
\caption{\label{Fig7}}
\end{figure}

\newpage

\begin{center}
\begin{tabular}{|c||c|c|c|c|}\hline
Ion                          &         $Ca^+$         &
$Hg^+$             &         $Ba^+$         &         $Yb^+$
      \\ \hline
level 0                      & $4s\, {}^2 S_{1/2}$    &
$5d^{10}6s^{2}\, {}^2 S_{1/2}$ &  $6s \,{}^2 S_{1/2}$   & $4f^{14}
6s^2\, {}^2 S_{1/2}$ \\ \hline
level 1                      & $3d\, {}^2 D_{5/2}$    &
$5d^{9}6s^{2}\, {}^2 D_{5/2}$  &  $5d \,{}^2 D_{5/2}$   & $4f^{13}
6s^2\, {}^2 F_{7/2}$ \\ \hline
level 2                      & $4s\, {}^2 P_{3/2}$    &
$5d^{10}6p^{2}\, {}^2 P_{1/2}$ &  $6s \,{}^2 P_{3/2}$   & $4f^{14}
6p^2\, {}^2 P_{3/2}$ \\ \hline\hline
$\;\omega_{01}\,[s^{-1}]\;$  & $2.61\cdot 10^{15}$    & $6.7\cdot
10^{15}$             &  $1.07\cdot 10^{15}$   &  $4.04\cdot 10^{15}$
 \\ \hline
$\;\omega_{02}\,[s^{-1}]\;$  & $4.76\cdot 10^{15}$    & $11.4\cdot
10^{15}$            &  $4.14\cdot 10^{15}$   &  $5.7\cdot 10^{15}$
\\ \hline\hline
$\;\Gamma_{out}\,[s^{-1}]\;$ & $4.95\cdot 10^6$ (H) &
	                       $1.5\cdot 10^8$  (H) &
                               $16.6\cdot 10^6$ (I) &
                               $0.9\cdot 10^6$	(F) \\ \hline
$\;\Gamma_{33\rightarrow 00}\,[s^{-1}]\;$ & $67.5\cdot 10^6$ (G) &
				            $5.26\cdot 10^8$ (H) &
					    $45.5\cdot 10^6$ (I) &
					    $60.2\cdot 10^6$ (F) \\ \hline\hline
$\;\;L(\eta=1)\;\;$          &      $16.0$            &
$15.0$             &  $21.0$                &  $32.0$
\\ \hline
$\;\;L(\eta=0.01)\;\;$       &      $3.7$             &
$3.9$              &  $5.1$                 &  $32.0$
\\ \hline
\end{tabular}
\end{center}
\vspace*{1.cm}
\begin{center}
	Plenio and Knight TABLE 3
\end{center}

\newpage

\begin{center}
\begin{tabular}{|c||c|c|c|c|}\hline
Ion                          &         $Ca^+$         &
$Hg^+$             &         $Ba^+$         &         $Yb^+$
      \\ \hline
level 0                      & $4s\, {}^2 S_{1/2}$    &
$5d^{10}6s^{2}\, {}^2 S_{1/2}$ &  $6s \,{}^2 S_{1/2}$   & $4f^{14}
6s^2\, {}^2 S_{1/2}$ \\ \hline
level 2                      & $3d\, {}^2 D_{5/2}$    &
$5d^{9}6s^{2}\, {}^2 D_{5/2}$  &  $5d \,{}^2 D_{5/2}$   & $4f^{13}
6s^2\, {}^2 F_{7/2}$ \\ \hline
level 3                      & $4s\, {}^2 P_{3/2}$    &
$5d^{10}6p^{2}\, {}^2 P_{1/2}$ &  $6s \,{}^2 P_{3/2}$   & $4f^{14}
6p^2\, {}^2 P_{3/2}$ \\ \hline\hline
$\;\omega_{02}\,[s^{-1}]\;$  & $2.61\cdot 10^{15}$    & $6.7\cdot
10^{15}$             &  $1.07\cdot 10^{15}$   &  $4.04\cdot 10^{15}$
 \\ \hline
$\;\omega_{13}\,[s^{-1}]\;$  & $4.76\cdot 10^{15}$    & $11.4\cdot
10^{15}$            &  $4.14\cdot 10^{15}$   &  $5.7\cdot 10^{15}$
\\ \hline\hline
$\;\Gamma_{out}\,[s^{-1}]\;$  & $4.95\cdot 10^6$ (J) &
			        $1.5\cdot 10^8$ (H)  &
 	                        $16.6\cdot 10^6$ (I) &
                                $0.9\cdot 10^6$	(F) \\ \hline\hline
$\;\Gamma_{33\rightarrow 00}\,[s^{-1}]\;$ & $67.5\cdot 10^6$ (G)&
					    $5.26\cdot 10^8$ (H)&
 					    $45.5\cdot 10^6$ (I)&
					    $60.2\cdot 10^6$ (F) \\ \hline
$\;\;L(\eta=1)\;\;$          &      $27.0$            &
$26$               &  $38.0$                &  $73.0$
\\ \hline
$\;\;L(\eta=0.01)\;\;$       &      $5.0$             &
$4.9$              &  $7.2$                 &  $73.0$
\\ \hline
\end{tabular}
\end{center}
\vspace*{1.cm}
\begin{center}
	Plenio and Knight TABLE 4
\end{center}


\begin{thebibliography}{99}
\harvarditem{Andersen \& S{\o}rensen}{1973}{Andersen1} T. Andersen
and G. S{\o}rensen,
1973
Systematic trends in atomic transition probabilities in neutral and
singly-ionized zinc,
cadmium and mercury, Journal of Quantitative
Spectroscopy and Radiative Transfer {\bf 13}, 369
%
\harvarditem{Augst {\em et al}}{1989}{Augst1}. Augst, S., Strickland,
D.,
 Meyerhofer, D.D., Chin, S.L. \& Eberly, J.H., 1989 Tunnel ionization
of noble-gases in
a high-intensity laser field, Phys. Rev. Lett. {\bf 63}, 2212-2215
and
references therein
%
\harvarditem{Barenco {\em et al}}{1995a}{Barenco1} Barenco, A.,
Deutsch, D.,
 Ekert, A. \& Josza, R., 1995 Conditional quantum dynamics and logic
gates,
Phys. Rev. Lett. {\bf 74}, 4083-4086
%
\harvarditem{Barenco {\em et al}}{1995b}{Barenco3} Barenco, A.,
Bennett, C.H.,
Cleve, R.,  DiVincenzo, D.P.,
 Margolus, N., Shor, P.,  Sleator, T., Smolin, J.A. \& Weinfurter,
H., 1995
Elementary gates for quantum computation, Phys. Rev. A
{\bf 52}, 3457-3467
%
\harvarditem{Barenco}{1996}{Barenco0} Barenco, A., Quantum physics
and
computers, Contemp. Phys. {\bf 37}, 375-395 (1996)
%
\harvarditem{Barenco {\em et al}}{1996}{Barenco2} Barenco, A.,
,Ekert, A.,
 Suominen, K.A. \& P. Torm{\"a}, 1996 Approximate quantum Fourier
transform,
Phys. Rev. A {\bf 54}, 139-146
%
\harvarditem{Bashkin \& Stoner}{1978}{Bashkin1} Bashkin, S., \&
Stoner, J.O.,
1978
{\em Atomic Energy-Level and
Grotian Diagrams} Vol II, Sulfur I to Titanium XXII p. 360-361,
(North Holland, Amsterdam)
%
\harvarditem{Bell {\em et al}}{1992}{Bell1} Bell, A.S., Gill, P.,
Klein, H.A.,
 Levick  A.P. \&  Rowley, W.R.C.,1992,  Precision measurement of the
${}^2F_{7/2} - {}^2D_{5/2}$ $3.43\mu m$ interval in trapped Yb+,
J. Mod. Opt. {\bf 39}, 381-387
%
\harvarditem{Bergquist {\em et al}}{1985}{Bergquist1}  Bergquist,
J.C., Wineland, D.J.,
Itano, W.I., Hemmati, H., Daniel, H.-U. \& Leuchs, G., 1985
Energy and Radiative Lifetime
of the $5d^9 6s^2 \,{}^2D_{5/2}$ state in Hg II by Doppler free two
photon laser spectroscopy,
Phys. Rev. Lett. {\bf 55}, 1567-1570
%
%
\harvarditem{Calderbank \& Shor}{1996}{Calderbank1} Calderbank, A.R.
\& Shor, P.W.,
1996 Good quantum error-correcting codes exist, Phys. Rev. A
{\bf 54}, 1098-1105
%
\harvarditem{Carmichael}{1993}{Carmichael1}
Carmichael, H.J., 1993, {\em An Open Systems
Approach to Quantum Optics}, Lecture Notes in Physics, (Springer,
Berlin).
%
\harvarditem{Cirac \& Zoller}{1995}{Cirac1}  Cirac, J.I. \& and
Zoller, P.,
1995 Quantum computation with cold trapped ions, Phys. Rev. Lett.
{\bf 74}, 4091-4094
%
\harvarditem{Cirac {\em et al}}{1996}{Cirac2} Cirac, J.I.,
Pellizzari, T. \& Zoller, P.,
1996 Enforcing coherent evolution in dissipative quantum dynamics,
submitted
to Science
%
\harvarditem{Dalibard {\em et al}}{1992}{Dalibard1}
Dalibard, J., Castin, Y. \& Molmer, K.,  1992
Wave-function approach to dissipative processes
in quantum optics, Phys. Rev. Lett. {\bf 68}, 580-583
%
\harvarditem{DiVincenzo \& Shor}{1996}{DiVincenzo1} DiVincenzo, D.P.
\& Shor, P.W.,
1996 Fault-Tolerant Error Correction with Efficient Quantum Codes, in
PRL {\bf 77},
September
%
\harvarditem{Ekert {\em et al}}{1996}{Ekert1} Ekert, A., \& Josza,
R., 1996
Rev. Mod. Phys. {\bf 68}, 733-753
%
\harvarditem{Ekert \& Machiavello}{1996}{Ekert2} Ekert, A., \&
Macchiavello, C.
1996 Quantum Error Correction for Communication, Phys. Rev. Lett.
{\bf 77}, 2585-2588
%
\harvarditem{Eriksen \& Poulsen}{1980}{Eriksen1} Eriksen, P. \&
Poulsen, O., 1980 Lifetime
measurements of the $6p^2P$ and $6d^2D$ levels in Hg(II), Journal of
Quantitative
Spectroscopy and Radiative Transfer {\bf 23}, 599
%
\harvarditem{Fawcett \& Wilson}{1991}{Fawcett1} Fawcett, B.C. \&
Wilson, 1991
 M., Computed oscillator strengths, Lande g values and lifetimes in
Yb+,
Atomic Data and Nuclear Data Tables {\bf 47}, 241-317
%
\harvarditem{Gallagher}{1967}{Gallagher1} Gallagher, A., 1967
Oscillator strengths of Ca II, Sr II and Ba II,
Phys. Rev. {\bf 157}, 24
%
\harvarditem{Garg}{1996}{Garg1} Garg, A., 1996 Decoherence in
ion-trap computers,
Phys. Rev. Lett. {\bf 77}, 964-967
%
\harvarditem{Gill {\em et al}}{1995}{Gill1} Gill, P., Klein, H.A.,
Roberts, M.,
Rowley, W.R.C. \& Taylor, P., 1995 Absolute measurement of the
${}^2S_{1/2} - {}^2D_{5/2}$ $411nm$ interval and the search for the
${}^2S_{1/2} - {}^2F_{572}$ $467nm$ transition in laser-cooled
trapped Yb+,
{\em Proc. 5th Symposium on Frequency Standards and Metrology}
(Woods Hole, MA)
%
\harvarditem{Gosselin {\em et al}}{1988}{Gosselin1} Gosselin, R.N.,
Pinnington, E.H.
\& Ansbacher, W., 1988 Measurement of the lifetime of the 4p levels
in CAII using laser excitation of a fast beam,
Phys. Rev. A {\bf 38}, 4887-4890
%
\harvarditem{Hegerfeldt \& Wilser}{1991}{Hegerfeldt1}
Hegerfeldt, G.C., and Wilser, T. S., 1991,
{\em Proceedings of the II. International Wigner Symposium}, edited
by
H. D. Doebner, W. Scherer, and F. Schroeck (World Scientific,
Singapore).
%
\harvarditem{Hughes {\em et al}}{1996}{Hughes1}
Hughes, R.J., James, D.F.V., Knill, E.H., Laflamme, R. \& Petschek,
A.G.,
1996,  Decoherence bounds on quantum computation with
trapped ions, Phys. Rev. Lett. {\bf 77}
%
\harvarditem{James \& Hughes}{1996}{James1} James, D.F.V. \& Hughes,
R.J.,
1996 Global limits on quantum computation with cold trapped ions I:
two level qubits, preprint
%
%
\harvarditem{Knight \& Garraway}{1996}{Knight1}
Knight, P.L. \& Garraway, B.M., 1996,
{\em Quantum Dynamics of Simple Systems. Proceedings of the
Forty Fourth Scottish Universities Summer School in Physics Stirling}
edited by G-L. Oppo, S.M. Barnett, E. Riis and M. Wilkinson
(Institute of Physics Publishing, Bristol)
%
\harvarditem{Knill \& Laflamme}{1996}{Knill1} Knill, E., \& Laflamme,
R.,
1996, A theory of quantum error-correcting code, lanl e-print
quant-ph/9604015
%
\harvarditem{Monroe {\em et al}}{1995}{Monroe1} Monroe, C., Meekhof,
D.M.,
King, B.E., Itano, W.M. \& Wineland, D.J.,  1995 Demonstration of a
fundamental
quantum logic gate, Phys. Rev. Lett {\bf 75}, 4714-4717
%
\bibitem{Palma1}  Palma, G.M.,  Suominen, K.-A. \& Ekert, A.K., 1996
Quantum computers and dissipation, Proc. Roy. Soc. {\bf A}452,
567-584
%
\harvarditem{Plenio \& Knight}{1996a}{Plenio2} Plenio, M.B. \&
Knight, P.L.,
1996 {\em Proceedings of the
2nd International Symposium on Fundamental Problems in Quantum
Physics},
edited by M. Ferrero and A. van der Merwe (Kluwer, Dordrecht)
%
\harvarditem{Plenio \& Knight}{1996b}{Plenio1} Plenio, M.B., and
Knight, P.L.,
1996 Realistic lower bounds for the factorization time of large
numbers on
a quantum computer, Phys. Rev. A {\bf 53}, 2986-2990
%
\harvarditem{Plenio \& Knight}{1996c}{Plenio4} Plenio, M.B., and
Knight, P.L.,
1996 The quantum jump approach to dissipative dynamics in quantum
optics, in
preparation for Rev. Mod. Phys.
%
\harvarditem{Plenio {\em et al}}{1996}{Plenio0} Plenio, M.B., Vedral,
V.
\& Knight, P.L., 1996 Computers and communication in the quantum
world,
Phys. World {\bf 9}, 19-20
%
\harvarditem{Plenio {\em et al}}{1996}{Plenio3} Plenio, M.B., Vedral,
V. \& Knight, P.L.,
1996 Conditional generation of error syndromes in fault-tolerant
error correction,
lanl e-print quant-ph/9608028
%
\harvarditem{Sauter {\em et al}}{1986a}{Sauter1} Sauter, Th., Blatt,
R., Neuhauser, W.
\& Toschek, P.E., 1986
Quantum Jumps observed in the fluorescence of a single ion,
Opt. Comm. {\bf 60}, 287-292
%
\harvarditem{Sauter {\em et al}}{1986b}{Sauter2}  Sauter, Th.,
Neuhauser, W., Blatt, R.
\&  Toschek, P.E., 1986
 Observation of quantum jumps,
Phys. Rev. Lett. {\bf 57}, 1696-1698
%
\harvarditem{Shor}{1994}{Shor1} Shor, P. W., 1994 in {\em Proceedings
of the 35th Annual
Symposium on the Foundations of Computer Science, Los Alamitos,}
CA (IEEE Computer Society Press, New York), p. 124
%
\harvarditem{Shor}{1995}{Shor2} Shor, P.W., 1995 Scheme for reducing
decoherence
in quantum computer memory, Phys. Rev. A {\bf 52}, R2493-2496
%
\harvarditem{Shor}{1996}{Shor3} Shor, P.W., 1996 Fault-Tolerant
Quantum Computation
lanl e-print quant-ph/9605011
%
\harvarditem{Sleator \& Weinfurter}{1995}{Sleator1} Sleator,T. \&
Weinfurter, H.,
1995 Realizable universal quantum logic gates, Phys. Rev. Lett.
{\bf 74}, 4087-4090
%
\harvarditem{Steane}{1996a}{Steane1} Steane, A.M., 1996
Error-correcting codes in
quantum-theory, Phys. Rev. Lett, {\bf 77}, 793-796
%
\harvarditem{Steane}{1996b}{Steane2} Steane, A.M., 1996 Multiple
Particle Interference
and Quantum Error Correction, to appear in Proc. R. Soc. Lond. A
%
\harvarditem{Vedral {\em et al}}{1996}{Vedral1} Vedral, V., Barenco,
A. \&
Ekert, A., 1996 Quantum networks for elementary arithmetic
operations, Phys. Rev. A
{\bf 54}, 147-153
%
\harvarditem{Wineland {\em et al}}{1992}{Wineland1} Wineland, D.J.,
Bollinger, J.J.,
Itano, W.M. \& Moore, F.L., 1992 Spin squeezing and reduced quantum
noise in
spectroscopy, Phys. Rev. A {\bf 46}, R6797-6800
%
\harvarditem{Wiese {\em et al}}{1969}{Wiese1} Wiese, W.L., Smith,
M.W. \& Miles, B.M.,
1969 {\em Atomic Transition Probabilitie, Vol II Sodium through
Calcium} p.251, (U.S.
Government Printing Office, Washington)
%
\end{thebibliography}
\end{document}